\documentclass[a4paper,12pt]{article}
\usepackage{amsmath}
\usepackage{amssymb}
\usepackage{amsfonts}
\usepackage{amsthm}
\usepackage{enumerate}
\usepackage[hyperindex,plainpages=false]{hyperref}
\usepackage[all]{xy}
\usepackage{graphicx}

\theoremstyle{plain}
\newtheorem{prop}{Proposition}[section]
\theoremstyle{definition}
\newtheorem{defi}{Definition}[section]

\numberwithin{equation}{section}

\newcounter{fcounter}

\newcounter{lcounter}

\newcounter{ccounter}

\newcounter{cpcounter}

\DeclareMathOperator{\tr}{tr}

\DeclareMathOperator{\Ad}{Ad}
\DeclareMathOperator{\sgn}{sgn}
\DeclareMathOperator{\Lie}{Lie}
\DeclareMathOperator{\Vect}{Vect}
\DeclareMathOperator{\Vol}{Vol}

\title{Observer dependent geometries}
\author{Manuel Hohmann\footnote{\href{mailto:manuel.hohmann@ut.ee}{manuel.hohmann@ut.ee}}\\Teoreetilise F\"u\"usika Labor, F\"u\"usika Instituut, Tartu \"Ulikool,\\Riia 142, 51014 Tartu, Estonia}

\begin{document}
\begin{titlepage}
\maketitle

\begin{abstract}
From general relativity we have learned the principles of general covariance and local Lorentz invariance, which follow from the fact that we consider observables as tensors on a spacetime manifold whose geometry is modeled by a Lorentzian metric. Approaches to quantum gravity, however, hint towards a breaking of these symmetries and the possible existence of more general, non-tensorial geometric structures. Possible implications of these approaches are non-tensorial transformation laws between different observers and an observer-dependent notion of geometry. In this work we review two different frameworks for observer dependent geometries, which may provide hints towards a quantization of gravity and possible explanations for so far unexplained phenomena: Finsler spacetimes and Cartan geometry on observer space. We discuss their definitions, properties and applications to observers, field theories and gravity.
\end{abstract}
\newpage
\tableofcontents
\end{titlepage}

\section{Geometry for observers and observables}\label{sec:introduction}
In order to establish a link to experiments, every physical theory needs to define the notions of observers and observables. From an experimentalist's point of view, an observation is the process of an observer performing an experiment in which he measures a number of physical quantities, called observables. Each measured observable is expressed by a single number or a set of numbers. In order to understand the meaning of these numbers from a theorist's point of view, and thus in a mathematical language, observers and observables must be modeled by mathematical objects, which can in turn be related to the outcomes of measurements. This model determines how the result of an observation depends on the observer who is performing it, and how the results obtained by different observers can be related to each other. In this work we will focus on geometric models for these relations.

We start our discussion from the viewpoint of general relativity. The most basic notion of general relativity is that of spacetime, which is modeled by a smooth manifold \(M\) equipped with a pseudo-Riemannian metric \(g\) of Lorentzian signature \((-,+,+,+)\), an orientation and a time orientation. Observers are modeled by world lines, which are smooth, future directed, timelike curves \(\gamma: \mathbb{R} \to M\). Their tangent vectors satisfy
\begin{equation}\label{eqn:timelike}
g_{ab}(\gamma(t))\dot{\gamma}^a(t)\dot{\gamma}^b(t) < 0\,.
\end{equation}
By a reparametrization we can always normalize the tangent vectors, so that
\begin{equation}\label{eqn:normalize}
g_{ab}(\gamma(t))\dot{\gamma}^a(t)\dot{\gamma}^b(t) = -1\,.
\end{equation}
In this case we call the curve parameter the proper time along the world line \(\gamma\) and denote it by the letter \(\tau\) instead of \(t\). The proper time along a timelike curve with arbitrary parametrization is given by the arc length integral
\begin{equation}\label{eqn:metricclock}
\tau_2 - \tau_1 = \int_{t_1}^{t_2}\sqrt{|g_{ab}(\gamma(t))\dot{\gamma}^a(t)\dot{\gamma}^b(t)|}dt\,.
\end{equation}
The clock postulate of general relativity states that any clock moving along the world line \(\gamma\) measures the proper time, independent of the construction of the clock. The prescription for the measurement of time is thus crucially linked to the Lorentzian metric of spacetime. Similarly, the metric provides a definition of rulers and the length of spacelike curves by the same expression~\eqref{eqn:metricclock} of the arc length integral. Finally, it also defines the angle \(\phi\) between two tangent vectors \(v, w \in T_xM\) at the same point \(x \in M\) as
\begin{equation}\label{eqn:metricangle}
\cos\phi = \frac{g_{ab}(x)v^aw^b}{\sqrt{g_{cd}(x)v^cv^d\,g_{ef}(x)w^ew^f}}\,.
\end{equation}
In summary, the Lorentzian metric \(g\) defines the \emph{geometry of spacetime}.

Closely related to the geometry of spacetime is the notion of causality. It answers the question which events on a spacetime manifold \(M\) can have a causal influence on which other events on \(M\). An event at \(x \in M\) can influence an event \(x' \in M\) if and only if there exists a continuous, future directed, causal (i.e., timelike or lightlike) curve from \(x\) to \(x'\). All events which can be influenced by \(x\) constitute the causal future of \(x\). Conversely, all events which can influence \(x'\) form the causal past of \(x'\). This structure, called the \emph{causal structure of spacetime}, is defined by the metric geometry via the definition of causal curves.

The Lorentzian spacetime metric serves several further purposes besides providing a definition of spacetime geometry and causality. We have already seen that it enters the definition of observer world lines as timelike curves, whose notion is thus also relevant when we consider the measurements of observables by these observers. Observables are modeled by tensor fields, which are smooth sections \(\Phi: M \to T^{r,s}M\) of a tensor bundle
\begin{equation}\label{eqn:tensorbundle}
T^{r,s}M = TM^{\otimes r} \otimes T^*M^{\otimes s}
\end{equation}
over \(M\). Their dynamics are consequently modeled by tensorial equations, which are derived from a diffeomorphism-invariant action of the generic form
\begin{equation}\label{eqn:metricmatter}
S_{\text{M}} = \int_Md^4x\sqrt{-g}\,\mathcal{L}(g,\Phi,\partial\Phi,\ldots)\,,
\end{equation}
where the Lagrange function \(\mathcal{L}\) depends on the metric geometry, the fields and their derivatives. Combining the notions of observers and observables we may define an observation by an observer with world line \(\gamma\) at proper time \(\tau\) as a measurement of the field \(\Phi(x)\) at the point \(x = \gamma(\tau)\). However, this definition yields us an element of the tensor space \(T^{r,s}_xM\), and not a set of numbers, as we initially presumed. We further need to choose a frame, by which we denote a basis \(f\) of the tangent space \(T_xM\). This frame allows us to express the tensor \(\Phi(x)\) in terms of its components with respect to \(f\). The tensor components of \(\Phi(x)\) are finally the numeric quantities which are measured in an experiment.

The frame \(f\) chosen by an observer to make measurements is usually not completely arbitrary. Since the basis vectors \(f_i\) are elements of the tangent space, they are characterized as being timelike, lightlike or spacelike and possess units of time or length. We can thus use the notions of time, length and angles defined by the spacetime metric to choose an orthonormal frame satisfying the condition
\begin{equation}\label{eqn:metriconframe}
g_{ab}f_i^af_j^b = \eta_{ij}
\end{equation}
with one unit timelike vector \(f_0\) and three unit spacelike vectors \(f_{\alpha}\). The clock postulate that proper time is measured by the arc length along the observer world line \(\gamma\) further implies a canonical choice of the timelike vector \(f_0\) as the tangent vector \(\dot{\gamma}(\tau)\) to the observer world line. This observer adapted orthonormal frame is a convenient choice for most measurements.

It follows immediately from this model of observables and observations how the measurements of the same observable made by two coincident observers, whose world lines \(\gamma\) and \(\gamma'\) meet at a common spacetime point \(x = \gamma(\tau) = \gamma'(\tau')\), must be translated between their frames of reference. If both observer frames \(f\) and \(f'\) are orthonormalized, the condition~\eqref{eqn:metriconframe} implies that they are related by a Lorentz transform \(\Lambda\). The same Lorentz transform must then be applied to the tensor components measured by one observer in order to obtain the tensor components measured by the other observer, using the standard formula
\begin{equation}\label{eqn:lorentzcov}
\Phi'^{a_1 \ldots a_r}{}_{b_1 \ldots b_s} = \Lambda^{a_1}{}_{c_1} \ldots \Lambda^{a_r}{}_{c_r}\Lambda^{d_1}{}_{b_1} \ldots \Lambda^{d_s}{}_{b_s}\Phi^{c_1 \ldots c_r}{}_{d_1 \ldots d_s}\,.
\end{equation}
This close connection between observations made using different observer frames constitutes the principle of local Lorentz invariance. It is a consequence of the fact that we model the geometry of spacetime, which in turn defines the notion of orthonormal frames, by a Lorentzian metric.

Even deeper implications arise from the fact that we model both observables and geometry by tensor fields on the spacetime manifold \(M\), and observations by measurements of tensor components. If we introduce coordinates on \(M\) and use their coordinate base in order to express the components of tensor fields, it immediately follows how these components translate under a change of coordinates. Moreover, since we model the dynamics of physical quantities by tensor equations, they are independent of any choice of coordinates. This coordinate freedom constitutes the principle of general covariance.

Besides its role in providing the background geometry which enters the definition of observers, observations and causality, the Lorentzian metric of spacetime has a physical interpretation on its own, being the field which carries the \emph{gravitational interaction}. It does not only govern the dynamics of matter fields, but is also influenced by their presence. This is reflected by the dynamics of gravity, which is governed by the Einstein-Hilbert action
\begin{equation}\label{eqn:ehaction}
S_{\text{EH}} = \frac{1}{2\kappa}\int_Md^4x\sqrt{-g}\,R\,,
\end{equation}
which together with the matter action~\eqref{eqn:metricmatter} yields the Einstein equations
\begin{equation}\label{eqn:metricgravity}
R_{ab} - \frac{1}{2}Rg_{ab} = \kappa T_{ab}\,.
\end{equation}
Understanding the geometry of spacetime as a dynamical quantity, which mutually interacts with matter fields, establishes a symmetric picture between both matter and gravity.

However, it is exactly this symmetry between gravity and matter which may lead us to new insights on the nature of spacetime geometry, and even question its description in terms of a Lorentzian metric, from which we derived a number of conclusions as stated above. This stems from the fact that all known matter fields in the standard model are nowadays described by quantum theories. While the process of quantization has been successfully applied to matter fields even beyond the standard model, it is significantly harder in the case of gravity. This difficulty has lead to a plethora of different approaches towards quantum gravity, many of which suggest modifications to the geometry of spacetime, or even resolve the unity of spacetime into a time evolution of spatial geometry. Main contenders which fall into this class are given by geometrodynamic theories such as loop quantum gravity~\cite{Ashtekar:1987gu,Thiemann:2007zz} and sum-over-histories formulations such as spin foam models~\cite{Rovelli:1995ac,Reisenberger:1996pu,Barrett:1997gw,Baez:1997zt} or causal dynamical triangulations~\cite{Ambjorn:1998xu,Ambjorn:2001cv,Ambjorn:2005qt}. Theories of this type introduce non-tensorial quantities, which may in turn suggest a breaking of general covariance at least at the quantum level. Moreover, other approaches to gravity may induce a breaking of local Lorentz invariance, for example, by a preferred class of observers, or test particles, described by a future unit timelike vector field~\cite{Brown:1994py,Jacobson:2000xp}.

The possible observer dependence of physical quantities beyond tensorial transformations motivates the introduction of spacetime geometries obeying a similar observer dependence, which generalize the well-known Lorentzian metric geometry. In this work we review and discuss two different, albeit similar, approaches to observer dependent geometries under the aspects of observers, causality and gravity. In section~\ref{sec:finsler} we review the concept of Finsler spacetimes~\cite{Pfeifer:2011tk,Pfeifer:2011xi,Pfeifer:2013gha}. We show that it naturally generalizes the causal structure of Lorentzian spacetimes, provides clear definitions of observers, observables and observations, serves as a background geometry for field theories and constitutes a model for gravity. In section~\ref{sec:cartan} we review the concept of observer space in terms of Cartan geometry~\cite{Gielen:2012fz}. Our discussion is based on the preceding discussion of Finsler spacetimes, from which we translate the notions of observers and gravity to Cartan language~\cite{Hohmann:2013fca}. We finally ponder the question which implications observer dependent geometries have on the nature of spacetime.

\section[Geometry of the clock postulate: Finsler spacetimes]{Geometry of the clock postulate:\\Finsler spacetimes}\label{sec:finsler}
As we have mentioned in the introduction, the metric geometry of spacetime serves multiple roles: it provides a causal structure, crucially enters the definition of observers, defines measures for length, time and angles and mediates the gravitational interaction. In this section we discuss a more general, non-metric spacetime geometry which is complete in the sense that it serves all of these roles. This generalized geometry is based on the concept of Finsler geometry~\cite{Chern,Bucataru}. Models of this type have been introduced as extensions to Einstein and string gravity~\cite{Horvath,Vacaru:2010fi,Vacaru:2007ng,Vacaru:2002kp}. In this work we employ the Finsler spacetime framework~\cite{Pfeifer:2011tk,Pfeifer:2011xi,Pfeifer:2013gha}, which is an extension of the well-known concept of Finsler geometry to Lorentzian signature, and review some of its properties and physical applications. This framework is of particular interest since, in addition to its aforementioned completeness, it can also be used to model small deviations from metric geometry and provides a possible explanation of the fly-by anomaly~\cite{Anderson:2008zza}.

\subsection{Definition of Finsler spacetimes}\label{subsec:finslerdef}
The starting point of our discussion is the clock postulate, which states that the time measured by an observer's clock moving along a timelike curve \(\gamma\) is the proper time \(\tau\) given by the arc length integral~\eqref{eqn:metricclock}. The expression
\begin{equation}\label{eqn:metricff}
F(\gamma(t),\dot{\gamma}(t)) = \sqrt{|g_{ab}(\gamma(t))\dot{\gamma}^a(t)\dot{\gamma}^b(t)|}
\end{equation}
under the integral depends on both the position \(\gamma(t)\) along the curve and the tangent vector \(\dot{\gamma}(t)\). Hence, it can be regarded as a function \(F: TM \to \mathbb{R}\) on the tangent bundle. The clock postulate thus states that the proper time measured by an observer's clock is given by the integral
\begin{equation}\label{eqn:finslerclock}
\tau_2 - \tau_1 = \int_{t_1}^{t_2}F(\gamma(t),\dot{\gamma}(t))dt\,,
\end{equation}
where \(F\) is the function on the tangent bundle given by equation~\eqref{eqn:metricff}.

For convenience we introduce a particular set \((x^a,y^a)\) of coordinates on \(TM\). Let \((x^a)\) be coordinates on \(M\). For \(y \in T_xM\) we then use the coordinates \((y^a)\) defined by
\begin{equation}
y = y^a\frac{\partial}{\partial x^a}\,.
\end{equation}
We call these coordinates induced by the coordinates \((x^a)\). As a further shorthand notation we use
\begin{equation}
\partial_a = \frac{\partial}{\partial x^a}\,, \qquad \bar{\partial}_a = \frac{\partial}{\partial y^a}
\end{equation}
for the coordinate basis of \(T_{(x,y)}TM\).

We now introduce a different, non-metric geometry of spacetime which still implements the clock postulate in the form of an arc length integral~\eqref{eqn:finslerclock}, but with a more general function \(F\) on the tangent bundle. Geometries of this type are known as Finsler geometries, and \(F\) is denoted the Finsler function. The choice of \(F\) we make here is not completely arbitrary. In order for the arc length integral to be well-defined and to obtain a suitable notion of spacetime geometry we need to preserve a few properties of the metric-induced Finsler function~\eqref{eqn:metricff}. In particular we will consider only Finsler functions which satisfy the following:

\begin{list}{\textrm{F\arabic{fcounter}}.}{\usecounter{fcounter}}
\item\label{finsler:fpositive}
\(F\) is non-negative, \(F(x,y) \geq 0\).
\item\label{finsler:fsmooth}
\(F\) is a continuous function on the tangent bundle \(TM\) and smooth where it is non-vanishing, i.e., on \(TM \setminus \{F = 0\}\).
\item\label{finsler:fhomorev}
\(F\) is positively homogeneous of degree one in the fiber coordinates and reversible, i.e.,
\begin{equation}\label{eqn:finslerhomo}
F(x,\lambda y) = |\lambda|F(x,y) \quad \forall \lambda \in \mathbb{R}\,.
\end{equation}
\end{list}

Property~\ref{finsler:fpositive} guarantees that the length of a curve is non-negative. We cannot demand strict positivity here, since already in the metric case we have the notion of lightlike curves \(\gamma\), for which \(F(\gamma(t),\dot{\gamma}(t)) = 0\). For the same reason of compatibility with the special case of a Lorentzian spacetime metric we cannot demand that \(F\) is smooth on all of \(TM\), since the metric Finsler function~\eqref{eqn:metricff} does not satisfy this condition. It does, however, satisfy the weaker condition~\ref{finsler:fsmooth}, which guarantees that the arc length integral depends smoothly on deformations of the curve \(\gamma\), unless these pass the critical region where \(F = 0\). Finally, we demand that the arc length integral is invariant under changes of the parametrization and on the direction in which the curve is traversed, which is guaranteed by condition~\ref{finsler:fhomorev}.

One may ask whether the Lorentzian metric \(g_{ab}\) can be recovered in case the Finsler function is given by~\eqref{eqn:metricff}. Indeed, the Finsler metric
\begin{equation}\label{eqn:fmetric}
g^F_{ab}(x,y) = \frac{1}{2}\bar{\partial}_a\bar{\partial}_bF^2(x,y)\,,
\end{equation}
which is defined everywhere on \(TM \setminus \{F = 0\}\), agrees with \(g_{ab}\) whenever \(y\) is spacelike and with \(-g_{ab}\) when \(y\) is timelike. However, for null vectors where \(F = 0\) we see that the Finsler metric \(g^F_{ab}\) is not well-defined, since for a general Finsler function \(F^2\) will not be differentiable. As a consequence any quantities derived from the metric, such as connections and curvatures, are not defined along the null structure, which renders this type of geometry useless for the description of lightlike geodesics. In the following we will therefore adopt the following definition of Finsler spacetimes which remedies this shortcoming~\cite{Pfeifer:2011tk}:

\begin{defi}[Finsler spacetime]
A \emph{Finsler spacetime} \((M,L,F)\) is a four-dimensional, connected, Hausdorff, paracompact, smooth manifold \(M\) equipped with continuous real functions \(L, F\) on the tangent bundle \(TM\) which has the following properties:
\begin{list}{\textrm{L\arabic{lcounter}}.}{\usecounter{lcounter}}
\item\label{finsler:lsmooth}
\(L\) is smooth on the tangent bundle without the zero section \(TM \setminus \{0\}\).
\item\label{finsler:lhomogeneous}
\(L\) is positively homogeneous of real degree \(n \geq 2\) with respect to the fiber coordinates of \(TM\),
\begin{equation}
L(x,\lambda y) = \lambda^nL(x,y) \quad \forall \lambda > 0\,,
\end{equation}
and defines the Finsler function \(F\) via \(F(x,y) = |L(x,y)|^{\frac{1}{n}}\).
\item\label{finsler:lreversible}
\(L\) is reversible: \(|L(x,-y)| = |L(x,y)|\).
\item\label{finsler:lhessian}
The Hessian
\begin{equation}
g^L_{ab}(x,y) = \frac{1}{2}\bar{\partial}_a\bar{\partial}_bL(x,y)
\end{equation}
of \(L\) with respect to the fiber coordinates is non-degenerate on \(TM \setminus X\), where \(X \subset TM\) has measure zero and does not contain the null set \(\{(x,y) \in TM | L(x,y) = 0\}\).
\item\label{finsler:timelike}
The unit timelike condition holds, i.e., for all \(x \in M\) the set
\begin{equation}
\Omega_x = \left\{y \in T_xM \left| |L(x,y)| = 1, g^L_{ab}(x,y) \text{ has signature } (\epsilon,-\epsilon,-\epsilon,-\epsilon)\right.\right\}
\end{equation}
with \(\epsilon = L(x,y)/|L(x,y)|\) contains a non-empty closed connected component \({S_x \subseteq \Omega_x \subset T_xM}\).
\end{list}
\end{defi}

One can show that the Finsler function \(F\) induced from the fundamental geometry function \(L\) defined above indeed satisfies the conditions~\ref{finsler:fpositive} to~\ref{finsler:fhomorev} we required. Further, the Finsler metric~\eqref{eqn:fmetric} is defined on \(TM \setminus \{L = 0\}\) and is non-degenerate on \(TM \setminus (X \cup \{L = 0\})\), where \(X\) is the degeneracy set of the Hessian \(g^L_{ab}\) defined in condition~\ref{finsler:lhessian} above. This definition in terms of the smooth fundamental geometry function \(L\) will be the basis of our discussion of Finsler spacetimes in the following sections, where we will see that it also extends the definitions of other geometrical structures such as connections and curvatures to the null structure.

\subsection{Causal structure and observers}\label{subsec:finslercausality}
\begin{figure}
\centerline{\includegraphics[width=72mm]{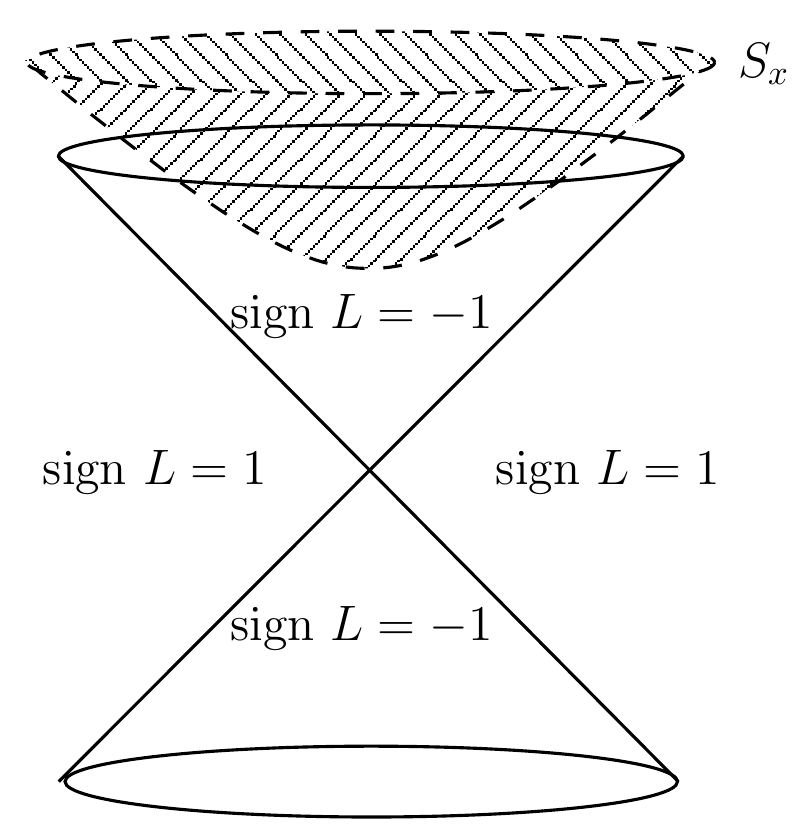}}
\caption{Light cone and future unit timelike vectors $S_x$ in the tangent space of a metric spacetime~\cite{Pfeifer:2011tk}.}
\label{fig:smcausal}
\end{figure}

The first aspect we discuss is the causal structure of Finsler spacetimes and the definition of observer trajectories. For this purpose we first examine the causal structure of metric spacetimes from the viewpoint of Finsler geometry, before we come to the general case. We have already mentioned in the introduction that the definition of causal curves is given by the split of the tangent spaces into timelike, spacelike and lightlike vectors. Figure~\ref{fig:smcausal} shows this split induced by the Lorentzian metric on the tangent space \(T_xM\). Solid lines mark the light cone which is constituted by null vectors. In terms of the fundamental geometry function
\begin{equation}
L(x,y) = g_{ab}(x)y^ay^b
\end{equation}
these are given by the condition \(L(x,y) = 0\). Outside the light cone we have spacelike vectors with \(L(x,y) > 0\), while inside the light cone we have timelike vectors with \(L(x,y) < 0\). The Hessian \(g^L_{ab} = g_{ab}\) therefore has the signature indicated in condition~\ref{finsler:timelike} inside the light cone. In both the future and the past light cones we find a closed subset with \(|L(x,y)| = 1\). Using the time orientation we pick one of these subsets and denote it the shell \(S_x\) of future unit timelike vectors.

\begin{figure}
\centerline{\includegraphics[width=132mm]{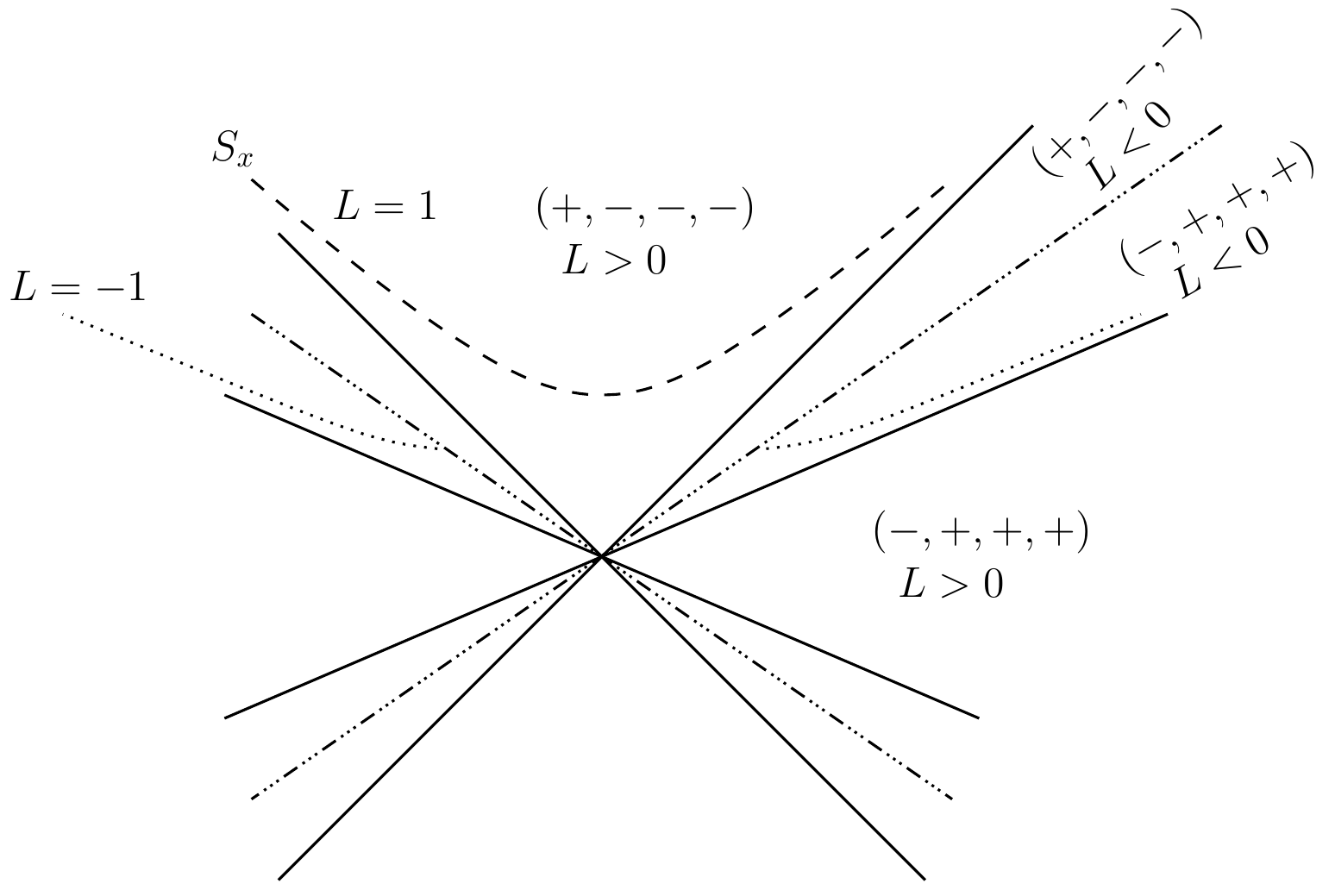}}
\caption{Null structure and future unit timelike vectors $S_x$ in the tangent space of a bimetric Finsler spacetime~\cite{Pfeifer:2011tk}.}
\label{fig:bmcausal}
\end{figure}

The shell \(S_x\) has the important property that rescaling yields a convex cone
\begin{equation}\label{eqn:convexcone}
C_x = \bigcup_{\lambda > 0}\lambda S_x \subset T_xM\,.
\end{equation}
The convexity of this cone is crucial for the interpretation of the elements of \(S_x\) as tangent vectors to observer world lines, as it is closely linked to the hyperbolicity of the dispersion relations of massive particles and the positivity of particle energies measured by an observer~\cite{Raetzel:2010je}. We require this property also for the future light cone of a Finsler spacetime. In order to find this structure in terms of the fundamental geometry function \(L\) consider the simple bimetric example
\begin{equation}
L(x,y) = h_{ab}(x)y^ay^bk_{cd}(x)y^cy^d
\end{equation}
with two Lorentzian metrics \(h_{ab}\) and \(k_{ab}\), where we assume that the light cone of \(k_{ab}\) lies in the interior of the light cone of \(h_{ab}\). The sign of \(L\) and the signature of \(g^L_{ab}\) on the tangent space \(T_xM\) are shown in figure~\ref{fig:bmcausal}. Solid lines mark the null structure \(L = 0\), while the dashed-dotted lines marks the degeneracy set \(X \cap T_xM\) of \(L\) as defined in condition~\ref{finsler:lhessian}. The remaining dashed and dotted lines mark the unit timelike vectors \(\Omega_x\) as defined in condition~\ref{finsler:timelike}; for these only the future directed tangent vectors are shown. The connected component marked by the dashed line is closed, while the one marked with the dotted line is not. Hence, the former marks the set \(S_x\). As the figure indicates, the set~\eqref{eqn:convexcone} indeed forms a convex cone for this simple bimetric example. It can be shown that condition~\ref{finsler:timelike} always implies the existence of a convex cone of observers~\cite{Pfeifer:2011tk}, in consistency with the requirement stated above.

It is now straightforward to define:

\begin{defi}[Observer world line]
A physical \emph{observer world line} on a Finsler spacetime is a curve \(\gamma: \mathbb{R} \to M\) such that at all times \(t\) the tangent vector \(\dot{\gamma}(t)\) lies inside the forward light cone \(C_{\gamma(t)}\), or in the unit timelike shell \(S_{\gamma(\tau)}\) if the curve parameter is given by the proper time \(\tau\).
\end{defi}

In the following section we will discuss which of these observers are further singled out by the Finsler spacetime geometry as being inertial observers.

\subsection{Dynamics for point masses}\label{subsec:finslerpm}
In the preceding section we have seen which trajectories are allowed for physical observers. We now turn our focus to a particular class of observers who follow the trajectories of freely falling test masses. These are denoted inertial observers, since in their local frame of reference gravitational effects can be neglected. On a metric spacetime they are given by those trajectories which extremize the arc lenth integral~\eqref{eqn:metricclock}. In Finsler geometry we can analogously obtain them from extremizing the proper time integral~\eqref{eqn:finslerclock}. Variation with respect to the curve yields the equation of motion
\begin{equation}\label{eqn:finslergeodesic}
\ddot{\gamma}^a + N^a{}_b(\gamma,\dot{\gamma})\dot{\gamma}^b = 0\,,
\end{equation}
where the coefficients \(N^a{}_b\) are given by the following definition:

\begin{defi}[Cartan non-linear connection]
The coefficients \(N^a{}_b\) of the \emph{Cartan non-linear connection} are given by
\begin{equation}\label{eqn:nonlincoeff}
N^a{}_b = \frac{1}{4}\bar{\partial}_b\left[g^{F\,ac}(y^d\partial_d\bar{\partial}_cF^2 - \partial_cF^2)\right]
\end{equation}
and define a connection in the sense that they induce a split of the tangent bundle over \(TM\),
\begin{equation}\label{eqn:ttmsplit}
TTM = HTM \oplus VTM\,,
\end{equation}
where \(HTM\) is spanned by \(\delta_a = \partial_a - N^b{}_a\bar{\partial}_b\) and \(VTM\) is spanned by \(\bar{\partial}_a\).
\end{defi}

In the case of a metric-induced Finsler function~\eqref{eqn:metricff} the coefficients \(N^a{}_b\) are given by
\begin{equation}\label{eqn:metricnlc}
N^a{}_b = \Gamma^a{}_{bc}y^c\,,
\end{equation}
where \(\Gamma^a{}_{bc}\) denotes the Christoffel symbols. The split~\eqref{eqn:ttmsplit} of \(TTM\) into horizontal and vertical subbundles plays an important role in Finsler geometry, as we will see in the following sections. For convenience we use the following adapted basis of \(TTM\):

\begin{defi}[Berwald basis]
The \emph{Berwald basis} is the basis
\begin{equation}\label{eqn:berwaldbasis}
\{\delta_a = \partial_a - N^b{}_a\bar{\partial}_b, \bar{\partial}_a\}
\end{equation}
of \(TTM\) which respects the split induced by the Cartan non-linear connection.
\end{defi}

For the dual basis we use the notation
\begin{equation}\label{eqn:dualberwaldbasis}
\{dx^a, \delta y^a = dy^a + N^a{}_bdx^b\}\,.
\end{equation}
It induces a similar split of the cotangent bundle \(T^*TM\) into the subbundles
\begin{equation}\label{eqn:tstmsplit}
T^*TM = H^*TM \oplus V^*TM\,.
\end{equation}
We can now reformulate the geodesic equation~\eqref{eqn:finslergeodesic} by making use of the geometry on \(TTM\). For this purpose we canonically lift the curve \(\gamma\) to a curve
\begin{equation}\label{eqn:canonicallift}
\Gamma = (\gamma,\dot{\gamma})
\end{equation}
in \(TTM\). The condition that \(\gamma\) is a Finsler geodesic then translates into the condition
\begin{equation}
\dot{\Gamma} = \dot{\gamma}^a\partial_a + \ddot{\gamma}^a\bar{\partial}_a = \dot{\gamma}^a\partial_a - \dot{\gamma}^bN^a{}_b\bar{\partial}_a = \dot{\gamma}^a\delta_a
\end{equation}
Since \(\dot{\gamma}^a\) is simply the tangent bundle coordinate \(y^a\), it thus follows that the canonical lift \(\Gamma\) of a Finsler geodesic must be an integral curve of the vector field which is defined as follows:

\begin{defi}[Geodesic spray]
The \emph{geodesic spray} \(\mathbf{S}\) is the vector field on \(TM\) which is defined by
\begin{equation}\label{eqn:geodspray}
\mathbf{S} = y^a\delta_a\,.
\end{equation}
\end{defi}

We now generalize this statement to null geodesics. Here we encounter two problems. First, we see that the coefficients~\eqref{eqn:nonlincoeff} of the non-linear connection are not well-defined for null vectors where \(F = 0\), since \(F\) is not differentiable on the null structure. We therefore need to rewrite their definition in terms of the fundamental geometry function \(L\). It turns out that it takes the same form
\begin{equation}\label{eqn:nonlincoeff2}
N^a{}_b = \frac{1}{4}\bar{\partial}_b\left[g^{L\,ac}(y^d\partial_d\bar{\partial}_cL - \partial_cL)\right]\,,
\end{equation}
where \(g^F\) has been replaced by \(g^L\) and \(F^2\) by \(L\). We can see that this is well-defined whenever \(g^L\) is non-degenerate, and thus in particular on the null structure. The second problem we encounter is that we derived the geodesic equation from extremizing the action~\eqref{eqn:finslerclock}, which vanishes identically in the case of null curves. We therefore need to use the constrained action
\begin{equation}\label{eqn:pmaction}
S[\gamma,\lambda] = \int_{t_1}^{t_2}\left(L(\gamma(t),\dot{\gamma}(t)) + \lambda(t)[L(\gamma(t),\dot{\gamma}(t)) - \kappa]\right)dt
\end{equation}
with a Lagrange multiplier \(\lambda\) and a constant \(\kappa\). A thorough analysis shows that the equations of motion derived from this action are equivalent to the geodesic equation~\eqref{eqn:finslergeodesic} also for null curves~\cite{Pfeifer:2011tk}.

The definitions of this and the preceding section provide us with the notions of general and inertial observers. In the following section we will discuss how these observers measure physical quantities and how the observations by different observers can be related.


\subsection{Observers and observations}\label{subsec:finslerlt}
As we have mentioned in the introduction, the notion of geometry in physics defines not only causality and the allowed trajectories of observers, but also their possible observations and the relation between observations made by different observers. In the case of metric spacetime geometry we have argued that observations are constituted by measurements of the components of tensor fields at a spacetime point \(x \in M\) with respect to a local frame \(f\) at \(x\). A particular class of frames singled out by the geometry and most convenient for measurements is given by the orthonormal frames. Different observations at the same spacetime point, but made with different local orthonormal frames, are related by Lorentz transforms. In this section we discuss a similar definition of observations on Finsler spacetimes and relate the observations made by different observers.

As a first step we need to generalize the notion of observables from metric spacetimes to Finsler spacetimes. In their definition in section~\ref{subsec:finslerdef} we have already seen that the geometry of Finsler spacetimes is defined by a homogeneous function \(L: TM \to \mathbb{R}\) on the tangent bundle, which in turn induces a Finsler function \(F\) and a Finsler metric \(g^F_{ab}\). These geometric objects explicitly depend not only on the manifold coordinates \(x^a\), but also on the coordinates \(y^a\) along the fibers of the tangent bundle \(TM\). It therefore appears natural that also observables should be functions not on the spacetime manifold, but homogeneous functions on its tangent bundle. A straightforward idea might thus be to model observables as homogeneous tensor fields over \(TM\), i.e., as sections of a tensor bundle
\begin{equation}
T^{r,s}TM = TTM^{\otimes r} \otimes T^*TM^{\otimes s}\,.
\end{equation}
However, since \(TM\) is an eight-dimensional manifold, each tensor index would then take eight values, so that the number of components of a tensor of rank \((r,s)\) would increase by a factor of \(2^{r + s}\). Since we do not observe these additional tensor components in nature, we will not follow this idea. Instead we define observables as tensor fields with respect to a different vector bundle over \(TM\), whose fibers are four-dimensional vector spaces generalizing the tangent spaces of \(M\).

In the preceding section we have seen that the Cartan non-linear connection~\eqref{eqn:nonlincoeff} of a Finsler spacetime equips the tangent bundle \(TTM\) of \(TM\) with a split~\eqref{eqn:ttmsplit} into a horizontal subbundle \(HTM\) and a vertical subbundle \(VTM\). The fibers of both subbundles are four-dimensional vector spaces. A particular section of \(HTM\), which we have already encountered and which is closely connected to Finsler geodesics, is the geodesic spray~\eqref{eqn:geodspray}. We therefore choose \(HTM\) as the bundle from which we define observables as follows:

\begin{defi}[Observable]
The \emph{observables} on a Finsler spacetime are modeled by homogeneous horizontal tensor fields, i.e., sections \(\Phi\) of the tensor bundle
\begin{equation}\label{eqn:htensorbundle}
H^{r,s}TM = HTM^{\otimes r} \otimes H^*TM^{\otimes s}
\end{equation}
over the tangent bundle \(TM\) of \(M\).
\end{defi}

Consequently we define observations in full analogy to the case of metric spacetime geometry:

\begin{defi}[Observation]
An \emph{observation} of an observable \(\Phi\) by an observer with world line \(\gamma\) at proper time \(\tau\) is a measurement of the components of the horizontal tensor \(\Phi(x,y)\) with respect to a basis \(f\) of the horizontal tangent space \(H_{(x,y)}TM\) at \(x = \gamma(\tau), y = \dot{\gamma}(\tau)\).
\end{defi}

As we have argued in the introduction, the most natural frame \(f\) an observer on a metric spacetime can choose is an orthonormal frame whose temporal component \(f_0\) agrees with his four-velocity \(\dot{\gamma}(\tau)\). If we wish to generalize this concept to Finsler spacetimes, we first need to map the basis vectors \(f_i\), which are now elements of \(HTM\), to \(TM\). For this purpose we use the differential \(\pi_*\) of the tangent bundle map \(\pi: TM \to M\),
\begin{equation}
\pi_*f_i = \pi_*(f_i^a\delta_a) = f_i^a\partial_a\,,
\end{equation}
which isomorphically maps every horizontal tangent space \(H_{(x,y)}TM\) to \(T_xM\). We can then orthonormalize the frame using the Finsler metric \(g^F_{ab}\), which now explicitly depends on the observer's four-velocity \(y = \pi_*f_0\). Taking into account the signature \((+,-,-,-)\) of the Finsler metric on timelike vectors inside the forward light cone we arrive at the following definition:

\begin{defi}[Orthonormal observer frame]
An \emph{orthonormal observer frame} on an observer world line \(\gamma\) at proper time \(\tau\) is a basis \(f\) of the horizontal tangent space \(H_{(x,y)}TM\) at \(x = \gamma(\tau), y = \dot{\gamma}(\tau)\) which has \(y = \pi_*f_0\) and is orthonormal with respect to the Finsler metric,
\begin{equation}\label{eqn:finsleronframe}
g^F_{ab}(x,y)f_i^af_j^b = -\eta_{ij}\,.
\end{equation}
\end{defi}

An important property of metric spacetimes is the fact that any two orthonormal observer frames \(f,f'\) at the same spacetime point \(x \in M\) are related by a unique Lorentz transform. Together with the definition that observations yield tensor components this property implies local Lorentz invariance, which means that the outcomes of measurements are related by the standard formula~\eqref{eqn:lorentzcov}. We now generalize this concept to Finsler spacetimes. For this purpose we consider two coincident observers whose world lines \(\gamma,\gamma'\) meet at \(x = \gamma(\tau) = \gamma'(\tau')\) together with orthonormal frames \(f,f'\) at \(x\). One immediately encounters the difficulty that \(f\) and \(f'\) are now bases of different vector spaces \(H_{(x,f_0)}TM\) and \(H_{(x,f'_0)}TM\). We therefore need to find a map between these vector spaces which in particular preserves the notion of orthonormality. The canonical map given by the isomorphisms \(\pi_*: H_{(x,f_0)}TM \to T_xM\) and \(\pi_*: H_{(x,f_0')}TM \to T_xM\), however, does not have this property. In the following we will therefore discuss a different map which will yield the desired generalization of Lorentz transformations.

In order to construct a map between the horizontal tangent spaces \(H_{(x,f_0)}TM\) and \(H_{(x,f'_0)}TM\) we employ the concept of parallel transport. We thus need a connection on the horizontal tangent bundle \(HTM\) with respect to which the Finsler metric is covariantly constant, so that the notion of orthonormality is preserved. In Finsler geometry an appropriate choice which satisfies these conditions is the Cartan linear connection on the tangent bundle \(TTM\), which is defined as follows:

\begin{defi}[Cartan linear connection]
The \emph{Cartan linear connection} \(\nabla\) is the connection on \(TTM\) defined by the covariant derivatives
\begin{equation}\label{eqn:clconnection}
\nabla_{\delta_a}\delta_b = F^c{}_{ab}\delta_c\,, \quad \nabla_{\delta_a}\bar{\partial}_b = F^c{}_{ab}\bar{\partial}_c\,, \quad \nabla_{\bar{\partial}_a}\delta_b = C^c{}_{ab}\delta_c\,, \quad \nabla_{\bar{\partial}_a}\bar{\partial}_b = C^c{}_{ab}\bar{\partial}_c\,,
\end{equation}
where the coefficients are given by
\begin{subequations}\label{eqn:clcoefficients}
\begin{align}
F^c{}_{ab} &= \frac{1}{2}g^{F\,cd}(\delta_ag^F_{bd} + \delta_bg^F_{ad} - \delta_dg^F_{ab})\,,\\
C^c{}_{ab} &= \frac{1}{2}g^{F\,cd}(\bar{\partial}_ag^F_{bd} + \bar{\partial}_bg^F_{ad} - \bar{\partial}_dg^F_{ab})\,.
\end{align}
\end{subequations}
\end{defi}

The Cartan linear connection is adapted to the Cartan non-linear connection~\eqref{eqn:nonlincoeff} in the sense that it respects the split~\eqref{eqn:ttmsplit} into horizontal and vertical components. By restriction it thus provides a connection on the horizontal tangent bundle. Given a curve \(v: [0,1] \to TM\) with \(v(0) = (x,f_0)\) and \(v(1) = (x,f'_0)\) we can then define a bijective map \(P_v\) from \(T_{(x,f_0)}TM\) to \(T_{(x,f'_0)}TM\) by parallel transport: it maps the vector \(w\) to \(P_vw = w'\), which is uniquely determined by the existence of a curve \(\hat{w}: [0,1] \to TTM\) satisfying
\begin{equation}
\hat{w}(s) \in T_{v(s)}TM\,, \quad \hat{w}(0) = w\,, \quad \hat{w}(1) = w'\,, \quad \nabla_{\dot{v}}w = 0\,.
\end{equation}
However, this map \(P_v\) in general depends on the choice of the curve \(v\). We therefore restrict ourselves to a particular class of curves. Note that \((x,f_0)\) and \((x,f'_0)\) have the same base point in \(M\), and are thus elements of the same fiber of the tangent bundle \(TM\). Hence it suffices to consider only curves which are fully contained in the same fiber. Curves of this type are vertical, i.e., their tangent vectors lie in the vertical tangent bundle \(VTM\). We further impose the condition that \(v\) is an autoparallel of the Cartan linear connection. This uniquely fixes the curve \(v\), provided that \(f'_0\) is in a sufficiently small neighborhood of \(f_0\).

Using the unique vertical autoparallel \(v\) defined above we can now generalize the notion of Lorentz transformations to coincident observers on a Finsler spacetime. Consider two observers meeting at \(x \in M\) and using frames \(f\) and \(f'\), i.e., orthonormal bases of \(H_{(x,f_0)}TM\) and \(H_{(x,f'_0)}TM\). The map \(P_v\) maps the horizontal basis vectors \(f_i\) to horizontal vectors \(P_vf_i\), which constitute a basis \(P_vf\) of \(H_{(x,f'_0)}TM\). Since \(f\) is orthonormal with respect to \(g^F_{ab}(x,f_0)\) and the Cartan linear connection preserves the Finsler metric, it follows that \(P_vf\) is orthonormal with respect to \(g^F_{ab}(x,f'_0)\). Since also \(f'\) is orthonormal with respect to the same metric, there exists a unique ordinary Lorentz transform mapping \(P_vf\) to \(f'\). The combination of the parallel transport along \(v\) and this unique Lorentz transform finally defines the desired generalized Lorentz transform.

The procedure to map bases of the horizontal tangent space between coincident observers further allows us to compare horizontal tensor components between these observers, so that they can communicate and compare their measurements of horizontal tensors. This corresponds to the transformation~\eqref{eqn:lorentzcov} of tensor components of observables between different observer frames in metric geometry. Since observables in metric geometry are modeled by spacetime tensor fields, their observation in one frame determines the measured tensor components in any other frame. This is not true on Finsler spacetimes, since we defined observables as fields on the tangent bundle \(TM\). They may therefore also possess a non-tensorial, explicit dependence on the four-velocity of the observer who measures them.

As in metric geometry, also in Finsler geometry the dynamics of tensor fields should be determined by a set of field equations which are derived from an action principle. This will be discussed in the next section.

\subsection{Field theory}\label{subsec:finslerft}
In the preceding section we have argued that observables on a Finsler spacetime are modeled by homogeneous horizontal tensor fields, which are homogeneous sections of the horizontal tensor bundle~\eqref{eqn:htensorbundle}. We will now discuss the dynamics of these observable fields. For this purpose we will use a suitable generalization of the action~\eqref{eqn:metricmatter} to horizontal tensor fields on a Finsler spacetime. This will be done in two steps. First we will lift the volume form from the spacetime manifold \(M\) to its tangent bundle \(TM\), then we generalize the Lagrange function \(\mathcal{L}\) to fields on a Finsler spacetime.

In order to define a volume form on \(TM\) we proceed in analogy to the volume form of metric geometry, which means that we choose the volume form \(\Vol_G\) of a suitable metric \(G\) on TM. We have already partly obtained this metric in the previous section when we discussed orthonormal observer frames. The definition of orthonormality we introduced corresponds to lifting the Finsler metric \(g^F_{ab}\) to a horizontal metric on \(TM\), which measures the length of horizontal vectors in \(HTM\). This metric needs to be complemented by a vertical metric, which analogously measures the length of vertical vectors in \(VTM\). Both metrics together constitute the desired metric on the tangent bundle. The canonical choice for this metric is given by the Sasaki metric defined as follows:

\begin{defi}[Sasaki metric]
The \emph{Sasaki metric} \(G\) is the metric on the tangent bundle \(TM\) which is defined by
\begin{equation}\label{eqn:sasakimetric}
G = -g^F_{ab}\,dx^a \otimes dx^b - \frac{g^F_{ab}}{F^2}\,\delta y^a \otimes \delta y^b\,.
\end{equation}
\end{defi}

The factor \(F^{-2}\) introduced here compensates for the intrinsic homogeneity of degree 1 of the one-forms \(\delta y^a\), so that the Sasaki metric is homogeneous of degree 0. This intrinsic homogeneity becomes clear from the definition~\eqref{eqn:dualberwaldbasis} of the dual Berwald basis, taking into account that the coefficients \(N^a{}_b\) are homogeneous of degree 1, as can be seen from their definition~\eqref{eqn:nonlincoeff}. Using the volume form \(\Vol_G\) of the Sasaki metric one can now integrate functions \(f\) on the tangent bundle,
\begin{equation}\label{eqn:tangbundint}
\int_{TM}\Vol_G f(x,y)\,.
\end{equation}
If one chooses the function \(f\) to be a suitable Lagrange function \(\mathcal{L}\) for a physical field \(\Phi\) on a Finsler spacetime, one encounters another difficulty. Since all geometric structures and matter fields \(\Phi\) are homogeneous, it is natural to demand the same from the Lagrange function. However, for a homogeneous function \(f\) the integral over the tangent bundle generically diverges, unless the function vanishes identically. This follows from the fact that along any ray \((x,\lambda y)\) with \(\lambda > 0\) in \(TM\) the value of \(f\) is given by \(\lambda^nf(x,y)\), where \(n\) is the degree of homogeneity. This difficulty can be overcome by integrating the function not over \(TM\), but over a smaller subset of \(TM\) which intersects each ray, which is not part of the null structure, exactly once, and which is defined as follows:

\begin{defi}[Unit tangent bundle]
The \emph{unit tangent bundle} of a Finsler spacetime is the set \(\Sigma \subset TM\) on which the Finsler function takes the value \(F = 1\).
\end{defi}

Note that \(\Sigma\) intersects each ray exactly once which is not part of the null structure. This suffices since the null structure is of measure 0 and therefore does not contribute to the integral~\eqref{eqn:tangbundint} over \(TM\). The canonical metric on \(\Sigma\) is given by the restriction
\begin{equation}\label{eqn:restrsasaki}
\tilde{G} = G|_{\Sigma}
\end{equation}
of the Sasaki metric, which finally determines the volume form \(\Vol_{\tilde{G}}\). This is the volume form we will use in the generalized action integral.

In the second part of our discussion we generalize the Lagrange function \(\mathcal{L}\) in the metric matter action~\eqref{eqn:metricmatter}. For simplicity we restrict ourselves here to $p$-form fields \(\Phi\) whose Lagrange function depends only on the field itself and its first derivatives \(d\Phi\). These are of particular interest since, e.g., the Klein-Gordon and Maxwell fields fall into this category. The most natural procedure to generalize the dynamics of a given field theory from metric to Finsler geometry is then to simply keep the formal structure of its Lagrange function \(\mathcal{L}\), but to replace the Lorentzian metric \(g\) by the Sasaki metric \(G\) and to promote the $p$-form field \(\Phi\) to a horizontal $p$-form field on \(TM\). The generalized Lagrange function we obtain from this procedure is now a function on \(TM\), which we can integrate over the subset \(\Sigma\) to form an action integral.

Using this procedure we encounter the problem that even though we have chosen \(\Phi\) to be horizontal, \(d\Phi\) will in general not be horizontal. In order to obtain consistent field equations we therefore need to modify our procedure. Instead of initially restricting ourselves to horizontal $p$-forms on the tangent bundle \(TM\), we let \(\Phi\) be an arbitrary $p$-form with both horizontal and vertical components. The purely horizontal components can then be obtained by applying the horizontal projector
\begin{equation}\label{eqn:horiproj}
P^H\Phi = \frac{1}{p!}dx^{a_1} \wedge \ldots \wedge dx^{a_p}\,\Phi(\delta_{a_1}, \ldots, \delta_{a_p})\,.
\end{equation}
In order to reduce the number of physical degrees of freedom to only these horizontal components we dynamically impose that the non-horizontal components vanish by introducing a suitable set of Lagrange multipliers \(\lambda\), so that the total action reads
\begin{equation}\label{eqn:matteraction}
S_{\text{M}} = \int_{\Sigma}\Vol_{\tilde{G}}\left[\mathcal{L}(G,\Phi,d\Phi) + \lambda(1 - P^H)\Phi\right]_{\Sigma}\,.
\end{equation}
Variation with respect to the Lagrange multipliers then yields the constraint that the vertical components of \(\Phi\) vanish. Variation with respect to these vertical components fixes the Lagrange multipliers. Finally, variation with respect to the horizontal components of \(\Phi\) yields the desired field equations. It can be shown that in the metric limit they reduce to the usual field equations derived from the action~\eqref{eqn:metricmatter} for matter fields on a metric spacetime~\cite{Pfeifer:2011xi}.

\subsection{Gravity}\label{subsec:finslergravity}
In the previous sections we have considered the geometry of Finsler spacetimes solely as a background geometry for observers, point masses and matter fields. We now turn our focus to the dynamics of Finsler geometry itself. As it is also the case for Lorentzian geometry, we will identify these dynamics with the dynamics of gravity. For this purpose we need to generalize the Einstein-Hilbert action, from which the gravitational field equations are derived, and the energy-momentum tensor, which acts as the source of gravity.

We start with a generalization of the Einstein-Hilbert action~\eqref{eqn:ehaction} to Finsler spacetimes. As in the case of matter field theories detailed in the preceding section this generalized action will be an integral not over spacetime \(M\), but over the unit tangent bundle \(\Sigma \subset TM\), since the geometry is defined in terms of the homogeneous fundamental geometry function \(L\) on TM. We have already seen that a suitable volume form on \(\Sigma\) is given by the volume form \(\Vol_{\tilde{G}}\) of the restricted Sasaki metric~\eqref{eqn:restrsasaki}. This leaves us with the task of generalizing the Ricci scalar \(R\) in terms on Finsler geometry.

The most natural and fundamental notion of curvature is defined by the Cartan non-linear connection~\eqref{eqn:nonlincoeff}, which we already encountered in the definition of Finsler geodesics in section~\ref{subsec:finslerpm} and which corresponds to the unique split~\eqref{eqn:ttmsplit} of the tangent bundle \(TTM\) into horizontal and vertical components. This split is also the basic ingredient for the following construction. The curvature of the Cartan non-linear connection measures the non-integrability of the horizontal distribution \(HTM\), i.e., the failure of the horizontal vector fields \(\delta_a\) to be horizontal. In fact their Lie brackets are vertical vector fields, which are used in the following definition:

\begin{defi}[Non-linear curvature]
The \emph{curvature of the non-linear connection} is the quantity \(R^c{}_{ab}\) which measures the non-integrability of the horizontal distribution induced by the Cartan non-linear connection,
\begin{equation}\label{eqn:nonlincurv}
[\delta_a,\delta_b] = (\delta_bN^c{}_a - \delta_aN^c{}_b)\bar{\partial}_c = R^c{}_{ab}\bar{\partial}_c\,.
\end{equation}
\end{defi}

The simplest scalar one can construct from the curvature coefficients defined by~\eqref{eqn:nonlincurv} is the contraction \(R^a{}_{ab}y^b\), so that the action for Finsler gravity takes the form
\begin{equation}\label{eqn:finsleraction}
S_{\text{F}} = \frac{1}{\kappa}\int_{\Sigma}\Vol_{\tilde{G}}R^a{}_{ab}y^b\,.
\end{equation}
In the case of a metric-induced Finsler function, in which the non-linear connection coefficients \(N^a{}_b\) are given by~\eqref{eqn:metricnlc}, the expression under the integral indeed reduces to the Ricci scalar, so that \(S_{\text{F}}\) is a direct generalization of the Einstein-Hilbert action~\eqref{eqn:ehaction}. In order to obtain a full gravitational theory this action needs to be complemented by a matter action, such as the field theory action~\eqref{eqn:matteraction} we encountered in the previous section. This total action then needs to be varied with respect to the mathematical object which fundamentally defines the spacetime geometry. On a Finsler spacetime this is the fundamental geometry function \(L\). Consequently, the gravitational field equations are not two-tensor equations as in general relativity, but instead the scalar equation
\begin{multline}\label{eqn:finslergravity}
\bigg[g^{F\,ab}\bar{\partial}_a\bar{\partial}_b(R^c{}_{cd}y^d) - 6\frac{R^a{}_{ab}y^b}{F^2}\\
+ 2g^{F\,ab}\left(\nabla_aS_b + S_aS_b + \bar{\partial}_a(y^c\delta_cS_b - N^c{}_bS_c)\right)\bigg]\bigg|_{\Sigma} = \kappa T|_{\Sigma}
\end{multline}
on the unit tangent bundle \(\Sigma\). Here \(T\) denotes the energy-momentum scalar obtained by variation of the matter action \(S_{\text{M}}\) with respect to the fundamental geometry function \(L\). For the field theory action~\eqref{eqn:matteraction} it is given by
\begin{equation}\label{eqn:finslerems}
T|_{\Sigma} = \left.\left\{\frac{nL}{\sqrt{-\tilde{G}}}\frac{\delta}{\delta L}\left[\sqrt{-\tilde{G}}\left(\mathcal{L}(G,\Phi,d\Phi) + \lambda(1 - P^H)\Phi\right)\right]\right\}\right|_{\Sigma}\,.
\end{equation}
It can be shown that in the metric limit the resulting gravitational field equation~\eqref{eqn:finslergravity} is equivalent to the Einstein equations~\eqref{eqn:metricgravity}, whose free indices are to be contracted with \(y^a\)~\cite{Pfeifer:2011xi}.

We finally remark that also the Cartan linear connection we used to define generalized Lorentz transformations in section~\ref{subsec:finslerlt} defines a notion of curvature, which may in principle be used to generalize the Einstein-Hilbert action. This curvature is defined as follows:

\begin{defi}[Linear curvature]
The \emph{curvature of the Cartan linear connection} is given by
\begin{equation}\label{eqn:clcurvature}
R(X,Y)Z = \nabla_X\nabla_YZ - \nabla_Y\nabla_XZ - \nabla_{[X,Y]}Z
\end{equation}
for vector fields \(X,Y,Z\) on \(TM\).
\end{defi}

Using the action~\eqref{eqn:clconnection} of the Cartan linear connection on the vector fields constituting the Berwald basis and the coefficients~\eqref{eqn:clcoefficients} one finds that its curvature can be written in the form
\begin{subequations}
\begin{align}
R(\delta_b,\delta_a)\delta_c &= R^d{}_{cab}\delta_d\,, & R(\delta_b,\delta_a)\bar{\partial}_c &= R^d{}_{cab}\bar{\partial}_d\,,\\
R(\bar{\partial}_b,\delta_a)\delta_c &= P^d{}_{cab}\delta_d\,, & R(\bar{\partial}_b,\delta_a)\bar{\partial}_c &= P^d{}_{cab}\bar{\partial}_d\,,\\
R(\bar{\partial}_b,\bar{\partial}_a)\delta_c &= S^d{}_{cab}\delta_d\,, & R(\bar{\partial}_b,\bar{\partial}_a)\bar{\partial}_c &= S^d{}_{cab}\bar{\partial}_d\,,
\end{align}
\end{subequations}
where the coefficients are given by
\begin{subequations}\label{eqn:clcurvcomp}
\begin{align}
R^d{}_{cab} &= \delta_bF^d{}_{ca} - \delta_aF^d{}_{cb} + F^e{}_{ca}F^d{}_{eb} - F^e{}_{cb}F^d{}_{ea} + C^d{}_{ce}(\delta_bN^e{}_a - \delta_aN^e{}_b)\,,\\
P^d{}_{cab} &= \bar{\partial}_bF^d{}_{ca} - \delta_aC^d{}_{cb} + F^e{}_{ca}C^d{}_{eb} - C^e{}_{cb}F^d{}_{ea} + C^d{}_{ce}\bar{\partial}_bN^e{}_a\,,\\
S^d{}_{cab} &= \bar{\partial}_bC^d{}_{ca} - \bar{\partial}_aC^d{}_{cb} + C^e{}_{ca}C^d{}_{eb} - C^e{}_{cb}C^d{}_{ea}\,.
\end{align}
\end{subequations}
In the metric limit the coefficient \(R^d{}_{cab}\) reduces to the Riemann tensor, while the remaining coefficients \(P^d{}_{cab}\) and \(S^d{}_{cab}\) vanish. One may therefore consider the term \(g^{F\,ab}R^c{}_{acb}\) as another generalization of the Ricci scalar to generate the gravitational dynamics on Finsler spacetimes. We do not pursue this idea further here and only remark that also other choices are possible.

\section[The local perspective: Cartan geometry of observer space]{The local perspective:\\Cartan geometry of observer space}\label{sec:cartan}
In the previous section we have seen that on Finsler spacetimes the definitions of observers and observables are promoted from geometrical structures on the spacetime manifold \(M\) to homogeneous geometrical structures on its tangent bundle \(TM\), and that this homogeneity fixes quantities on \(TM\) when they are given on the unit tangent bundle \(\Sigma\). We have also seen that measurements by an observer probe these structures along a lifted world line \(\Gamma = (\gamma,\dot{\gamma})\) in \(TM\). However, it follows from the definition of physical observer trajectories that every curve \(\Gamma\) is entirely confined to future unit timelike vectors, so that observations can be performed only on a smaller subset \(O \subset \Sigma\), which we denote observer space. In this section we will therefore restrict our discussion to observer space and equip it with a suitable geometrical structure in terms of Cartan geometry~\cite{Cartan,Sharpe}, which we derive from the previously defined Finsler geometry~\cite{Hohmann:2013fca}. While Cartan geometry turns out to be useful already as a geometry for spacetime in the context of gravity~\cite{Wise:2006sm}, it becomes even more interesting as a geometry for observer space~\cite{Gielen:2012fz} and provides a better insight into the role of Lorentz symmetry in canonical quantum gravity~\cite{Gielen:2011mk,Gielen:2012pn}.

\subsection{Definition of observer space}\label{subsec:observerspace}
We start our discussion with the definition of observer space as the space of all tangent vectors to a Finsler spacetime which are allowed as tangent vectors of normalized observer trajectories, i.e., observer trajectories which are parametrized by their proper time. This leads us to the definition:

\begin{defi}[Observer space]
The \emph{observer space} \(O\) of a Finsler spacetime \((M,L,F)\) is the set of all future unit timelike vectors, i.e., the union
\begin{equation}\label{eqn:observerspace}
O = \bigcup_{x \in M}S_x
\end{equation}
of all unit shells inside the forward light cones.
\end{defi}

Note that \(O\) is a seven-dimensional submanifold of \(TM\) and that its tangent spaces \(T_{(x,y)}O\) are spanned by the vectors \(v \in T_{(x,y)}TM\) which satisfy \(vF = 0\). Further, there exists a canonical projection \(\pi': O \to M\) onto the underlying spacetime manifold. The natural question arises which geometrical structure the Finsler geometry on the spacetime manifold \(M\) induces on its observer space \(O\). The structure which is most obvious already from our findings in the previous section is the restricted Sasaki metric \(\tilde{G}\), which we defined in~\eqref{eqn:restrsasaki} as the restriction of the full Sasaki metric \(G\) to \(\Sigma\) and which we now view as a metric on the smaller set \(O \subset \Sigma\). It follows from the signature of \(G\) that \(\tilde{G}\) has Lorentzian signature \((-,+,+,+,+,+,+)\).

Another structure which we already encountered in the previous section is the geodesic spray~\eqref{eqn:geodspray}. Since it preserves the Finsler function, \(\mathbf{S}F = 0\), it is tangent to the level sets of \(F\), and thus in particular tangent to observer space \(O\). It therefore restricts to a vector field on \(O\), which we denote the Reeb vector field:

\begin{defi}[Reeb vector field]
The \emph{Reeb vector field} \(\mathbf{r}\) is the restriction of the geodesic spray \(\mathbf{S}\) to \(O\),
\begin{equation}\label{eqn:reebvector}
\mathbf{r} = \left.y^a\delta_a\right|_O\,.
\end{equation}
\end{defi}

We now have a metric and a vector field on \(O\). Combining these two structures we can form the dual one-form \(\alpha\) of the Reeb vector field with respect to the restricted Sasaki metric \(\tilde{G}\), which we denote the contact form:

\begin{defi}[Contact form]
The \emph{contact form} is the dual one-form of the Reeb vector field \(\mathbf{r}\) with respect to the restricted Sasaki metric \(\tilde{G}\),
\begin{equation}\label{eqn:contactform}
\alpha = -\tilde{G}(\mathbf{r},.) = \left.g^F_{ab}y^a\,dx^b\right|_O = \left.\frac{1}{2}\bar{\partial}_aF^2\,dx^a\right|_O\,.
\end{equation}
\end{defi}

Conversely, the Reeb vector field is the unique vector field on \(O\) which is normalized by \(\alpha\) and whose flow preserves \(\alpha\), i.e., which satisfies
\begin{equation}\label{eqn:reebprops}
\mathcal{L}_{\mathbf{r}}\alpha = 0 \quad \text{and} \quad \alpha(\mathbf{r}) = 1\,.
\end{equation}
The naming of \(\alpha\) and \(\mathbf{r}\) originates from the notion of contact geometry. In this context a contact form on a $(2n + 1)$-dimensional manifold is defined as a one-form \(\alpha\), which is maximally non-integrable in the sense that the $(2n + 1)$-form \(\alpha \wedge d\alpha \wedge \ldots \wedge d\alpha\) is nowhere vanishing, hence defines a volume form, and the Reeb vector field is the unique vector field \(\mathbf{r}\) satisfying~\eqref{eqn:reebprops}. Indeed it turns out that the volume form defined by \(\alpha\) is simply the volume form of the Sasaki metric \(\tilde{G}\) on \(O\).

As we have seen in section~\ref{subsec:finslerpm} the Finsler geometry induces a split~\eqref{eqn:ttmsplit} of the eight-dimensional tangent bundle \(TTM\) into two four-dimensional subbundles \(VTM\) and \(HTM\), denoted the vertical and horizontal subbundles, respectively. A similar split also applies to the tangent bundle \(TO\) of observer space. It splits into the three subbundles
\begin{equation}\label{eqn:tangbundsplit}
TO = VO \oplus HO = VO \oplus \vec{H}O \oplus H^0O\,,
\end{equation}
which we denote the vertical, spatial and temporal subbundles, respectively. The vertical bundle \(VO\) is defined in analogy to the vertical tangent bundle \(VTM\) as the kernel of the differential \(\pi'_*\) of the canonical projection \(\pi': O \to M\). It is constituted by the tangent spaces to the shells \(S_x\) of unit timelike vectors at \(x \in M\) and hence three-dimensional. Its orthogonal complement with respect to the Sasaki metric \(\tilde{G}\) is the four-dimensional horizontal bundle \(HO\). One can easily see that the contact form \(\alpha\) vanishes on \(VO\). Its kernel on \(HO\) defines the three-dimensional spatial bundle \(\vec{H}O\). Finally, the orthogonal complement of \(\vec{H}O\) in \(HO\) is the one-dimensional temporal bundle \(H^0O\), which is spanned by the Reeb vector field \(\mathbf{r}\).

The split of the tangent bundle \(TO\) has a clear physical interpretation. Vertical vectors in \(VO\) correspond to infinitesimal generalized Lorentz boosts, which change the velocity of an observer, but not his position. They are complemented by horizontal vectors in \(HO\), which change the observer's position, but not his direction of motion. These further split into spatial translations in \(\vec{H}O\) and temporal translations in \(H^0O\) with respect to the observer's local frame. This interpretation will become clear when we discuss the split of the tangent bundle from a deeper geometric perspective using the language of Cartan geometry. We will give a brief introduction to Cartan geometry in the following section.

\subsection{Introduction to Cartan geometry}\label{subsec:cartanintro}
In order to describe the geometry of observer space, we make use of a framework originally developed by Cartan under the name ``method of moving frames''~\cite{Cartan}. His description of the geometry of a manifold \(M\) is based on a comparison to the geometry of a suitable model space. The latter is taken to be a homogeneous space, i.e., the coset space \(G/H\) of a Lie group \(G\) and a closed subgroup \(H \subset G\). Homogeneous spaces were extensively studied in Klein's Erlangen program and are hence also known as Klein geometries. Cartan's construction makes use of the fact that they carry the structure of a principal $H$-bundle \(\pi: G \to G/H\) and a connection given by the Maurer-Cartan one-form \(A \in \Omega^1(G,\mathfrak{g})\) on \(G\) taking values in the Lie algebra \(\mathfrak{g}\) of \(G\). Using these structures in order to describe the local geometry of \(M\), a Cartan geometry is defined as follows:

\begin{defi}[Cartan geometry]
Let \(G\) be a Lie group and \(H \subset G\) a closed subgroup of \(G\). A \emph{Cartan geometry} modeled on the homogeneous space \(G/H\) is a principal $H$-bundle \(\pi: P \to M\) together with a $\mathfrak{g}$-valued one-form \(A \in \Omega^1(P,\mathfrak{g})\), called the \emph{Cartan connection} on \(P\), such that
\begin{list}{\textrm{C\arabic{ccounter}}.}{\usecounter{ccounter}}
\item\label{cartan:isomorphism}
For each \(p \in P\), \(A_p: T_pP \to \mathfrak{g}\) is a linear isomorphism.
\item\label{cartan:equivariant}
\(A\) is $H$-equivariant: \((R_h)^*A = \mathrm{Ad}(h^{-1}) \circ A\) \(\forall h \in H\).
\item\label{cartan:mcform}
\(A\) restricts to the Maurer-Cartan form on vertical vectors \(v \in \ker\pi_*\).
\end{list}
\end{defi}
Instead of describing the Cartan geometry in terms of the Cartan connection \(A\), which is equivalent to specifying a linear isomorphism \(A_p: T_pP \to \mathfrak{g}\) for all \(p \in P\) due to condition~\ref{cartan:isomorphism}, we can use the inverse maps \(\underline{A}_p = A_p^{-1}: \mathfrak{g} \to T_pP\). For each \(a \in \mathfrak{g}\) they define a section \(\underline{A}(a)\) of the tangent bundle, which we denote a fundamental vector field:
\begin{defi}[Fundamental vector fields]
Let \((\pi: P \to M, A)\) be a Cartan geometry modeled on \(G/H\). For each \(a \in \mathfrak{g}\) the \emph{fundamental vector field} \(\underline{A}(a)\) is the unique vector field such that \(A(\underline{A}(a)) = a\).
\end{defi}
We can therefore equivalently define a Cartan geometry in terms of its fundamental vector fields, due to the following proposition:
\begin{prop}
Let \((\pi: P \to M, A)\) be a Cartan geometry modeled on \(G/H\) and \(\underline{A}: \mathfrak{g} \to \Vect P\) its fundamental vector fields. Then the properties~\ref{cartan:isomorphism} to~\ref{cartan:mcform} of \(A\) are respectively equivalent to the following properties of \(\underline{A}\):
\begin{list}{\textrm{C\arabic{cpcounter}}'.}{\usecounter{cpcounter}}
\item\label{cartan:isomorphism2}
For each \(p \in P\), \(\underline{A}_p: \mathfrak{g} \to T_pP\) is a linear isomorphism.
\item\label{cartan:equivariant2}
\(\underline{A}\) is $H$-equivariant: \(R_{h*} \circ \underline{A} = \underline{A} \circ \mathrm{Ad}(h^{-1})\) \(\forall h \in H\).
\item\label{cartan:canonical}
\(\underline{A}\) restricts to the canonical vector fields on \(\mathfrak{h}\).
\end{list}
\end{prop}

We illustrate these definitions using a physically motivated example. Let \(\tilde{\pi}: P \to M\) be the oriented, time-oriented, orthonormal frame bundle of a Lorentzian manifold \((M,g)\). It carries the structure of a principal $H$-bundle, where \(H = \mathrm{SO}_0(3,1)\) is the proper orthochronous Lorentz group. The homogeneous space \(G/H\) can be any of the maximally symmetric de Sitter, Minkowski or anti-de Sitter spacetimes, which is achieved by choosing the group \(G\) to be
\begin{equation}\label{eqn:groups}
G = \begin{cases}
\mathrm{SO}_0(4,1) & \text{for } \Lambda > 0 \Leftrightarrow \text{de Sitter spacetime,}\\
\mathrm{ISO}_0(3,1) & \text{for } \Lambda = 0 \Leftrightarrow \text{Minkowski spacetime,}\\
\mathrm{SO}_0(3,2) & \text{for } \Lambda < 0 \Leftrightarrow \text{anti-de Sitter spacetime,}
\end{cases}
\end{equation}
where \(\mathrm{ISO}_0(3,1) = \mathrm{SO}_0(3,1) \ltimes \mathbb{R}^{3,1}\) is the proper orthochronous Poincar\'{e} group and the subscript~\(0\) indicates the connected component of the corresponding group. Here \(\Lambda\) denotes the cosmological constant on the respective maximally symmetric spacetime and does not necessarily agree with the physical cosmological constant.

We further need to equip the frame bundle \(\tilde{\pi}: P \to M\) with a Cartan connection. For this purpose we introduce a component notation for elements of the Lie algebra \(\mathfrak{g} = \Lie G\) and its subalgebras. First observe that \(\mathfrak{g}\) splits into irreducible subrepresentations of the adjoint representation of \(H \subset G\),
\begin{equation}\label{eqn:algsplit1}
\mathfrak{g} = \mathfrak{h} \oplus \mathfrak{z}\,.
\end{equation}
These subspaces correspond to infinitesimal Lorentz transforms \(\mathfrak{h} = \Lie H\) and infinitesimal translations \(\mathfrak{z} \cong \mathfrak{g}/\mathfrak{h}\) of the homogeneous spacetimes \(G/H\). We can use this split to uniquely decompose any algebra element \(a \in \mathfrak{g}\) in the form
\begin{equation}\label{eqn:component}
a = h + z = \frac{1}{2}h^i{}_j\mathcal{H}_i{}^j + z^i\mathcal{Z}_i\,,
\end{equation}
where \(\mathcal{H}_i{}^j\) are the generators of \(\mathfrak{h} = \mathfrak{so}(3,1)\) and \(\mathcal{Z}_i\) are the generators of translations on \(G/H\). They satisfy the algebra relations
\begin{gather}
[\mathcal{H}_i{}^j,\mathcal{H}_k{}^l] = \delta^j_k\mathcal{H}_i{}^l - \delta^l_i\mathcal{H}_k{}^j + \eta_{ik}\eta^{lm}\mathcal{H}_m{}^j - \eta^{jl}\eta_{km}\mathcal{H}_i{}^m\,,\label{eqn:algebra}\\
[\mathcal{H}_i{}^j,\mathcal{Z}_k] = \delta^j_k\mathcal{Z}_i - \eta_{ik}\eta^{jl}\mathcal{Z}_l\,, \qquad [\mathcal{Z}_i,\mathcal{Z}_j] = \sgn\Lambda\,\eta_{ik}\mathcal{H}_j{}^k\,.\nonumber
\end{gather}
The last expression explicitly depends on the choice of the group \(G\), which can conveniently be expressed using the sign of the cosmological constant \(\Lambda\).

We can now apply this component notation to the Cartan connection \(A\). We first split \(A = \omega + e\) into a $\mathfrak{h}$-valued part \(\omega\) and a $\mathfrak{z}$-valued part \(e\). The latter we set equal to the solder form, which in component notation can be written as
\begin{equation}\label{eqn:solderform}
e^i = f^{-1}{}^i_adx^a\,,
\end{equation}
where the coordinates \((f_i^a)\) on the fibers of \(P\) are defined as the components of the frames \(f_i\) in the coordinate basis of the manifold coordinates \((x^a)\), and \(f^{-1}{}^i_a\) denote the corresponding inverse frame components. For the $\mathfrak{h}$-valued part \(\omega\) we choose the Levi-Civita connection. Given a curve \(\tau \mapsto (x(\tau),f(\tau))\) on \(P\) it measures the covariant derivative of the frame vectors \(f_i\) along the projected curve \(\tau \mapsto x(\tau)\) on \(M\). For a tangent vector \(v \in TP\) this yields
\begin{equation}\label{eqn:covderivative}
f_j\omega^j{}_i(v) = \nabla_{\tilde{\pi}_*(v)}f_i\,.
\end{equation}
Using the same component notation as above it reads
\begin{equation}\label{eqn:levicivita}
\omega^j{}_i = f^{-1}{}^j_adf_i^a + f^{-1}{}^j_af_i^b\Gamma^a{}_{bc}dx^c\,,
\end{equation}
where \(\Gamma^a{}_{bc}\) denotes the Christoffel symbols. It is not difficult to check that the $\mathfrak{g}$-valued one-form~\(A\) defined above indeed satisfies conditions~\ref{cartan:isomorphism} to~\ref{cartan:mcform} of a Cartan connection, and thus defines a Cartan geometry modeled on \(G/H\). Equivalently, we can describe the Cartan geometry in terms of the fundamental vector fields. Using the notation~\eqref{eqn:component} they take the form
\begin{equation}
\underline{A}(a) = h^i{}_jf_i^a\bar{\partial}^j_a + z^if_i^a\left(\partial_a - f_j^b\Gamma^c{}_{ab}\bar{\partial}^j_c\right)\,,
\end{equation}
where we have introduced the notation
\begin{equation}
\partial_a = \frac{\partial}{\partial x^a}\,, \quad \bar{\partial}^i_a = \frac{\partial}{\partial f_i^a}
\end{equation}
for tangent vectors to the frame bundle \(P\). A well-known result of Cartan geometry states that the metric \(g\) can be reconstructed from the Cartan connection, up to a global scale factor.

We finally remark that the Cartan geometry provides a split of the tangent bundle \(TP\) which has a similar physical interpretation as the split~\eqref{eqn:tangbundsplit} of \(TO\). This split is induced by the decomposition~\eqref{eqn:algsplit1} of the Lie algebra \(\mathfrak{g}\), which is carried over to the tangent spaces \(T_pP\) by the isomorphic mappings \(A_p\) as shown in the following diagram:
\begin{equation}\label{eqn:tangbundsplit1}
\xymatrix{V_pP \ar@{^(->>}[d]_{\omega} \ar@{}[r]|{\oplus} \ar@{}[dr]|{+} & H_pP \ar@{^(->>}[d]_e \ar@{}[r]|{=} \ar@{}[dr]|{=} & T_pP \ar@{^(->>}[d]_A\\
\mathfrak{h} \ar@{}[r]|{\oplus} & \mathfrak{z} \ar@{}[r]|{=} & \mathfrak{g}}
\end{equation}
The vertical subbundle \(VP\) is constituted by the tangent spaces to the fibers of the bundle \(\tilde{\pi}: P \to M\), which are given by the kernel of the differential \(\tilde{\pi}_*\) of the canonical projection. This is a direct consequence of condition~\ref{cartan:mcform} on the Cartan connection. The elements of \(VP\) can be viewed as infinitesimal local Lorentz transformations, which change only the local frame \(f\) and leave the base point \(x\) unchanged. Conversely, the elements of the horizontal subbundle \(HP\) correspond to infinitesimal translations, which change the base point \(x\) without changing the orientation of the local frame \(f\). This follows from the fact that we constructed the $\mathfrak{h}$-valued part \(\omega\) of the Cartan connection from the Levi-Civita connection.

\subsection{Cartan geometry of observer space}\label{subsec:cartanobs}
We will now employ Cartan geometry in order to describe the geometry of observer space. Hereby we will proceed in analogy to the metric spacetime example discussed in the previous section, where we constructed a Cartan connection on the orthonormal frame bundle. For this purpose we refer to the definition of orthonormal observer frames in section~\ref{subsec:finslerlt}. If we translate this definition to the context of observer space geometry, we find that an observer frame at \((x,y) \in O\) is a basis of the horizontal tangent space \(H_{(x,y)}O\) such that \(\pi'_*f_0 = y\) and the normalization~\eqref{eqn:finsleronframe} holds. Equivalently, we can make use of the differential \(\pi'_*\) of the canonical projection \(\pi': O \to M\), which isomorphically maps \(H_{(x,y)}O\) to \(T_xM\), and regard frames as bases of \(T_xM\), in analogy to the case of metric geometry. Here we choose the latter and define:

\begin{defi}[Observer frames]
The space \(P\) of \emph{observer frames} of a Finsler spacetime \((M,L,F)\) with observer space \(O\) is the space of all oriented, time-oriented tangent space bases \(f\) of \(M\), such that the basis vector \(f_0\) lies in \(O\) and the frame is orthonormal with respect to the Finsler metric,
\begin{equation}
g^F_{ab}(x,f_0)f_i^af_j^b = -\eta_{ij}\,.
\end{equation}
\end{defi}

One can now easily see that although there exists a canonical projection \(\tilde{\pi}: P \to M\), which assigns to an observer frame its base point on \(M\), it does in general not define a principal $H$-bundle, where \(H\) is the Lorentz group as in the preceding section. This follows from the fact that the generalized Lorentz transforms discussed in section~\ref{subsec:finslerlt} do not form a group, but only a grupoid. However, this is not an obstruction, as it is our aim to construct a Cartan geometry on \(O\) and not on \(M\). Indeed the projection \(\pi: P \to O\), which simply discards the spatial frame components, carries the structure of a principal $K$-bundle, where by \(K\) we denote the rotation group \(\mathrm{SO}(3)\). It acts on \(P\) by rotating the spatial frame components. The Cartan geometry on observer space will thus be modeled on the homogeneous space \(G/K\) instead of \(G/H\).

We further need to equip \(\pi: P \to O\) with a Cartan connection which generalizes the Cartan connection on the metric frame bundle displayed in the previous section. Here we can proceed in full analogy and choose as the $\mathfrak{z}$-valued part \(e\) of the connection the solder form. The expression in component notation,
\begin{equation}\label{eqn:connectionz}
e^i = f^{-1}{}^i_adx^a\,,
\end{equation}
agrees with the analogous expression~\eqref{eqn:solderform} in metric geometry. For the $\mathfrak{h}$-valued part \(\omega\) we generalize the Levi-Civita connection~\eqref{eqn:levicivita}. Recall from section~\ref{subsec:finslerlt} that the tangent space \(TM\) of a Finsler spacetime, and hence also its observer space \(O \subset TM\), is equipped with the Cartan linear connection~\eqref{eqn:clconnection}. We can therefore replace the projection \(\tilde{\pi}\) to \(M\) in~\eqref{eqn:covderivative} with the projection \(\pi\) to \(O\) and define
\begin{equation}\label{eqn:covderivative2}
f_j\omega^j{}_i(v) = \nabla_{\pi_*(v)}f_i\,,
\end{equation}
where \(\nabla\) now denotes the Cartan linear connection. In component notation this yields the expression
\begin{align}
\omega^j{}_i &= f^{-1}{}^j_adf_i^a + f^{-1}{}^j_af_i^b\left[F^a{}_{bc}dx^c + C^a{}_{bc}(N^c{}_ddx^d + df_0^c)\right]\nonumber\\
&= \frac{1}{2}\left(\delta_i^k\delta_l^j - \eta^{jk}\eta_{il}\right)f^{-1}{}^l_a\,df_k^a + \frac{1}{2}\eta^{jk}f_i^bf_k^c(\delta_bg^F_{ac} - \delta_cg^F_{ab})dx^a\,,\label{eqn:connectionh}
\end{align}
where the coefficients \(C^a{}_{bc}\) and \(F^a{}_{bc}\) are the coefficients of the Cartan linear connection~\eqref{eqn:clcoefficients}. From the Cartan connection~\eqref{eqn:connectionz} and~\eqref{eqn:connectionh} we then find the fundamental vector fields
\begin{subequations}\label{eqn:fundvecfields}
\begin{align}
\underline{A}(h) &= \left(h^i{}_jf_i^a - h^i{}_0f^b_if^c_jC^a{}_{bc}\right)\bar{\partial}^j_a\,,\label{eqn:vecfieldh}\\
\underline{A}(z) &= z^if_i^a\left(\partial_a - f_j^bF^c{}_{ab}\bar{\partial}^j_c\right)\label{eqn:vecfieldz}
\end{align}
\end{subequations}
for \(h \in \mathfrak{h}\) and \(z \in \mathfrak{z}\). One easily checks that indeed \(\underline{A}_p = A_p^{-1}\) for all \(p \in P\), so that condition~\ref{cartan:isomorphism} is satisfied. Another simple calculation shows that also conditions~\ref{cartan:equivariant} and~\ref{cartan:mcform} are satisfied, so that \(A\) defines a Cartan geometry.

The Cartan geometry on the observer frame bundle \(\pi: P \to O\) induces a split of the tangent bundle \(TP\) in analogy the split~\eqref{eqn:tangbundsplit1} we observed for the Cartan geometry of a metric spacetime. Since the observer space Cartan geometry is modeled on \(G/K\) instead of \(G/H\) we first decompose the Lie algebra \(\mathfrak{g}\) into irreducible subrepresentations of the adjoint representation of \(K \subset G\),
\begin{equation}\label{eqn:algsplit2}
\mathfrak{g} = \mathfrak{k} \oplus \mathfrak{y} \oplus \vec{\mathfrak{z}} \oplus \mathfrak{z}^0\,.
\end{equation}
The subspaces we encounter here are the rotation algebra \(\mathfrak{k} = \Lie K\), the rotation-free Lorentz boosts \(\mathfrak{y} \cong \mathfrak{h}/\mathfrak{k}\), as well as the spatial and temporal translations \(\mathfrak{z} = \vec{\mathfrak{z}} \oplus \mathfrak{z}^0\) of the homogeneous spacetimes. We can decompose the Cartan connection accordingly and obtain the following split of the tangent spaces \(T_pP\):
\begin{equation}\label{eqn:tangbundsplit2}
\xymatrix{R_pP \ar@{^(->>}[d]_{\Omega} \ar@{}[r]|{\oplus} \ar@{}[dr]|{+} & B_pP \ar@{^(->>}[d]_b \ar@{}[r]|{\oplus} \ar@{}[dr]|{+} & \vec{H}_pP \ar@{^(->>}[d]_{\vec{e}} \ar@{}[r]|{\oplus} \ar@{}[dr]|{+} & H^0_pP \ar@{^(->>}[d]_{e^0} \ar@{}[r]|{=} \ar@{}[dr]|{=} & T_pP \ar@{^(->>}[d]_A\\
\mathfrak{k} \ar@{}[r]|{\oplus} & \mathfrak{y} \ar@{}[r]|{\oplus} & \vec{\mathfrak{z}} \ar@{}[r]|{\oplus} & \mathfrak{z}^0 \ar@{}[r]|{=} & \mathfrak{g}}
\end{equation}
The elements of these subbundles correspond to infinitesimal rotations of observer frames in \(RP\), infinitesimal rotation-free Lorentz boosts in \(BP\) as well as translations along the spatial and temporal frame directions in \(\vec{H}P\) and \(H^0P\), respectively. For convenience we introduce a component notation for the algebra-valued one-forms \(\Omega \in \Omega^1(P,\mathfrak{k})\), \(b \in \Omega^1(P,\mathfrak{y})\), \(\vec{e} \in \Omega^1(P,\vec{\mathfrak{z}})\) and \(e^0 \in \Omega^1(P,\mathfrak{z}^0)\) in the form
\begin{equation}\label{eqn:conncomponents}
A = \Omega^{\alpha}\mathcal{R}_{\alpha} + b^{\alpha}\mathcal{L}_{\alpha} + e^{\alpha}\mathcal{Z}_{\alpha} + e^0\mathcal{Z}_0\,,
\end{equation}
where \(\mathcal{R}_{\alpha}, \mathcal{L}_{\alpha}, \mathcal{Z}_{\alpha}, \mathcal{Z}_0\) are the generators of rotations, Lorentz boosts as well as spatial and temporal translations. The ten components \(\Omega^{\alpha}, b^{\alpha}, e^{\alpha}, e^0\) are ordinary one-forms on \(P\). Note that for each \(p \in P\) they are linearly independent and thus constitute a basis of \(T^*_pP\). In a similar fashion we will write the fundamental vector fields \(\underline{A}\) in the decomposed form
\begin{equation}\label{eqn:vectcomponents}
\underline{A}(r^{\alpha}\mathcal{R}_{\alpha} + l^{\alpha}\mathcal{L}_{\alpha} + z^{\alpha}\mathcal{Z}_{\alpha} + z^0\mathcal{Z}_0) = r^{\alpha}\underline{\Omega}_{\alpha} + l^{\alpha}\underline{b}_{\alpha} + z^{\alpha}\underline{e}_{\alpha} + z^0\underline{e}_0\,,
\end{equation}
where the ten components \(\underline{\Omega}_{\alpha}, \underline{b}_{\alpha}, \underline{e}_{\alpha}, \underline{e}_0\) are ordinary vector fields on \(P\). They constitute bases of the tangent spaces \(T_pP\) which respect the split into the respective subspaces \(R_pP, B_pP, \vec{H}_pP, H^0_pP\) and are dual to the aforementioned cotangent space bases.

Recall from section~\ref{subsec:observerspace} that the tangent bundle \(TO\) of observer space features a split~\eqref{eqn:tangbundsplit} into Lorentz boosts and spatial and temporal translations which is similar to the split~\eqref{eqn:tangbundsplit2}. In fact these two splits are closely related. For each frame \(p \in P\) the differential \(\pi_*\) of the bundle projection isomorphically maps the subspaces of \(T_pP\), except the kernel \(R_pP\), to the corresponding subspaces of \(T_{\pi(p)}O\), as shown in the following diagram:
\begin{equation}\label{eqn:tangbundmap}
\xymatrix{R_pP \ar[d]_{\pi_*} \ar@{}[r]|{\oplus} & B_pP \ar@{^(->>}[d]_{\pi_*} \ar@{}[r]|{\oplus} & \vec{H}_pP \ar@{^(->>}[d]_{\pi_*} \ar@{}[r]|{\oplus} & H^0_pP \ar@{^(->>}[d]_{\pi_*} \ar@{}[r]|{=} & T_pP \ar[d]_{\pi_*}\\
0 & V_{\pi(p)}O \ar@{}[r]|{\oplus} & \vec{H}_{\pi(p)}O \ar@{}[r]|{\oplus} & H^0_{\pi(p)}O \ar@{}[r]|{=} & T_{\pi(p)}O}
\end{equation}
We see that we obtain the split of \(TO\), which we previously derived directly from Finsler geometry, also by using Cartan geometry. This observation brings us to the question of whether the observer space Cartan geometry also yields us the geometric structures on observer space we defined in section~\ref{subsec:observerspace} -- the Sasaki metric, the contact form and the Reeb vector field.

In order to relate geometric objects on \(O\) to the Cartan connection \(A\) and the fundamental vector fields \(\underline{A}\) on \(P\), one naturally makes use of the bundle projection \(\pi: P \to O\). Its pushforward \(\pi_*\) maps tangent vectors on \(P\) to tangent vectors on \(O\), as displayed also in diagram~\eqref{eqn:tangbundmap}. However, since \(\pi\) is not injective, and thus fails to be a diffeomorphism, it does not allow us to carry vector fields or differential forms from \(P\) to \(O\). We therefore need to enhance the relation between these spaces with a section \(s: O \to P\). It allows us evaluate the fundamental vector fields \(\underline{A}(a)\) for \(a \in \mathfrak{g}\) on the image of \(s\) and apply the differential \(\pi_*\), which yields us vector fields
\begin{equation}
\underline{\tilde{A}}(a) = \pi_* \circ \underline{A}(a) \circ s
\end{equation}
on \(O\). Note that these depend on the choice of the section \(s\). Using the component notation~\eqref{eqn:vectcomponents} we can define component vector fields on \(O\) by
\begin{equation}\label{eqn:vectfieldproj}
\underline{\tilde{\Omega}}_{\alpha} = \pi_* \circ \underline{\Omega}_{\alpha} \circ s\,, \quad \underline{\tilde{b}}_{\alpha} = \pi_* \circ \underline{b}_{\alpha} \circ s\,, \quad \underline{\tilde{e}}_{\alpha} = \pi_* \circ \underline{e}_{\alpha} \circ s\,, \quad \underline{\tilde{e}}_0 = \pi_* \circ \underline{e}_0 \circ s\,.
\end{equation}
It follows from~\eqref{eqn:tangbundmap} that \(\underline{\tilde{\Omega}}_{\alpha}\) vanishes, since the vector fields \(\underline{\Omega}_{\alpha}\) lie inside the rotation subbundle \(RP\) and thus in the kernel of \(\pi_*\). Further we find that the remaining vector fields \(\underline{b}_{\alpha}, \underline{e}_{\alpha}, \underline{e}_0\) constitute bases of the subspaces \(V_oO, \vec{H}_oO, H^0_oO\) of \(T_oO\) for each \(o \in O\). This shows that the fundamental vector fields \(\underline{\tilde{A}}_o\) evaluated at \(o\) isomorphically map the vector space \(\mathfrak{y} \oplus \vec{\mathfrak{z}} \oplus \mathfrak{z}^0\) to \(T_oO\) while respecting the split into subspaces. The inverse maps \(\tilde{A}_o = \underline{\tilde{A}}_o^{-1}\) therefore constitute a one-form
\begin{equation}
\tilde{A} = \tilde{b}^{\alpha}\mathcal{L}_{\alpha} + \tilde{e}^{\alpha}\mathcal{Z}_{\alpha} + \tilde{e}^0\mathcal{Z}_0 \in \Omega^1(O, \mathfrak{y} \oplus \vec{\mathfrak{z}} \oplus \mathfrak{z}^0)\,,
\end{equation}
whose components are the pullbacks of the components \(b^{\alpha}, e^{\alpha}, e^0\) on the image of the section \(s\).

Since the one-form \(\tilde{A}\) and fundamental vector fields \(\underline{\tilde{A}}\) defined above depend on the choice of the section, we now pose the question how they are related if we choose different sections \(s\) and \(s'\). Recall that \(\pi: P \to O\) is a principal $K$-bundle, so that any two sections are related by a local gauge transform, i.e., by a function \(k: O \to K\). Under this gauge transform the fundamental vector fields transform as
\begin{equation}
\underline{\tilde{A}}'(a) = \pi_* \circ \underline{A}(a) \circ R_k \circ s = \pi_* \circ \underline{A}(\Ad(k)(a)) \circ s = \left(\underline{\tilde{A}} \circ \Ad(k)\right)(a)
\end{equation}
using the irreducible subrepresentations of the adjoint representation of \(K\) on \(\mathfrak{g}\). Similarly, the one-forms transform as
\begin{equation}
\tilde{A}' = \Ad(k^{-1}) \circ \tilde{A}
\end{equation}
Since the adjoint representation of \(K\) acts trivially on the subspace \(\mathfrak{z}^0\) it immediately follows that the component fields \(\tilde{e}^0\) and \(\underline{\tilde{e}}_0\) are independent of the choice of the section \(s\). From the expressions~\eqref{eqn:connectionh} and~\eqref{eqn:connectionz} of the Cartan connection and the fundamental vector fields~\eqref{eqn:fundvecfields} in terms of Finsler geometry we see that these are simply the contact form~\eqref{eqn:contactform} and the Reeb vector field~\eqref{eqn:reebvector},
\begin{equation}
\tilde{e}^0 = \alpha\,, \quad \underline{\tilde{e}}_0 = \mathbf{r}\,.
\end{equation}
We have thus expressed these structures on \(O\) in terms of the Cartan connection on \(P\). It further turns out that the Sasaki metric takes the form
\begin{equation}\label{eqn:cartansasaki}
\tilde{G} = \eta_{ij}\,\tilde{e}^i \otimes \tilde{e}^j + \delta_{\alpha\beta}\,\tilde{b}^{\alpha} \otimes \tilde{b}^{\beta}\,,
\end{equation}
and is thus also expressed in terms of the Cartan connection. Note that also this is invariant under changes of the section, which act as a local rotation of the component fields. The same applies to its volume form
\begin{equation}\label{eqn:cartansasakivol}
\Vol_{\tilde{G}} = \epsilon_{ijkl}\epsilon_{\alpha\beta\gamma}\tilde{e}^i \wedge \tilde{e}^j \wedge \tilde{e}^k \wedge \tilde{e}^l \wedge \tilde{b}^{\alpha} \wedge \tilde{b}^{\beta} \wedge \tilde{b}^{\gamma}\,.
\end{equation}
In the following sections we will make use of these structures are their expressions in terms of Cartan geometry in order to provide definitions for observers and observations in analogy to those given in section~\ref{sec:finsler} using Finsler geometry.

\subsection{Observers and observations}\label{subsec:cartanpm}
We now come to the description of observers and their measurements in the language of Cartan geometry on observer space. In the following we will discuss which curves on observer space correspond to the trajectories of physical observers. In particular we will define the notion of inertial observers using elements of Cartan geometry.

In section~\ref{subsec:finslercausality} we have discussed the notion of physical observers on a Finsler spacetime. We have defined the trajectories of physical observers as those curves \(\tau \mapsto \gamma(\tau)\) on a Finsler spacetime, whose tangent vectors \(\dot{\gamma}(\tau)\) in arc length parametrization lie in the future unit timelike shell \(S_{\gamma(\tau)} \subset T_{\gamma(\tau)}M\). If we lift these curves canonically to curves \(\tau \mapsto (\gamma(\tau),\dot{\gamma}(\tau))\) on \(TM\), we thus see that they are entirely contained in observer space \(O \subset TM\). This leads to a very simple definition of physical trajectories on observer space:

\begin{defi}[Observer trajectory]
A physical \emph{observer trajectory} is a curve \(\Gamma\) on observer space which is the canonical lift \(\Gamma = (\gamma,\dot{\gamma})\)  of an observer world line \(\gamma\) on the underlying Finsler spacetime.
\end{defi}

We will now rewrite this condition in terms of Cartan geometry. First observe that canonical lifts in \(O\) are exactly those curves \(\Gamma\) such that the tangent vector of the projected curve \(\pi' \circ \Gamma\) in \(M\) reproduces \(\Gamma\),
\begin{equation}
\Gamma(\tau) = \frac{d}{d\tau}\pi'(\Gamma(\tau)) = \pi'_*\left(\dot{\Gamma}(\tau)\right)\,.
\end{equation}
One can easily see that this condition does not restrict the vertical components of \(\dot{\Gamma}(\tau)\), which lie inside the kernel \(VO\) of \(\pi'_*\) according to the split~\eqref{eqn:tangbundsplit}, and fully determines its horizontal components as a function of the position \(\Gamma(\tau)\) in observer space. It therefore defines a horizontal vector field \(\mathbf{h}\) on \(O\), i.e. a section \(\mathbf{h}: O \to HO\) of the horizontal tangent bundle which has the property that \(\pi'_* \circ \mathbf{h}: O \to TM\) is the identity on \(O\). The unique vector field which satisfies this condition is the Reeb vector field \(\mathbf{r} = \underline{\tilde{e}}_0\) defined in~\eqref{eqn:reebvector}. Hence, observer trajectories are those curves \(\Gamma\) on \(O\) whose horizontal tangent vector components are given by the Reeb vector field. We can further rewrite this condition by introducing the projectors
\begin{equation}\label{eqn:tangbundproj}
P_V = \underline{\tilde{b}}_{\alpha} \otimes \tilde{b}^{\alpha}\,, \quad P_{\vec{H}} = \underline{\tilde{e}}_{\alpha} \otimes \tilde{e}^{\alpha}\,, \quad P_{H^0} = \underline{\tilde{e}}_0 \otimes \tilde{e}^0\,, \quad P_H = P_{\vec{H}} + P_{H^0}
\end{equation}
onto the subbundles of \(TO\) and obtain the form \(P_H\dot{\Gamma}(\tau) = \mathbf{r}(\Gamma(\tau))\). Finally, inserting the explicit formulas for \(P_H\) and \(\mathbf{r}\) we arrive at the reformulated definition:

\begin{defi}[Observer trajectory]
A physical \emph{observer trajectory} is a curve \(\Gamma\) on observer space whose horizontal components are given by the Reeb vector field, i.e., which satisfies
\begin{equation}
\tilde{e}^i\dot{\Gamma}(\tau) = \delta^i_0\,.
\end{equation}
\end{defi}

A particular class of observers is given by inertial observers, whose trajectories follow those of freely falling test masses. In section~\ref{subsec:finslerpm} we have seen that these are given by Finsler geodesics, or equivalently by curves whose complete lift \((\gamma,\dot{\gamma})\) in \(TM\) is an integral curve of the geodesic spray~\eqref{eqn:geodspray}. We have further seen that the geodesic spray is tangent to observer space \(O \subset TM\) and defined the Reeb vector field \(\mathbf{r}\) as its restriction to observer space. It thus immediately follows that inertial observer trajectories on \(O\) are simply the integral curves of the Reeb vector field. Comparing this finding with the aforementioned definition we see that inertial observer trajectories are exactly those observer trajectories whose vertical tangent vector components vanish. We thus define, using only Cartan geometry:

\begin{defi}[Inertial observer trajectory]
An \emph{inertial observer trajectory} is an integral curve of the Reeb vector field, i.e., a curve \(\Gamma\) on observer space which satisfies
\begin{equation}
\tilde{b}^{\alpha}\dot{\Gamma}(\tau) = 0\,, \quad \tilde{e}^i\dot{\Gamma}(\tau) = \delta^i_0\,.
\end{equation}
\end{defi}

It appears now straightforward to translate the notions of observables and observations from Finsler geometry to Cartan geometry on observer space. A direct translation yields observables as sections of a horizontal tensor bundle, which is constructed from the horizontal subbundle \(HO\) in analogy to the horizontal tensor bundle \(H^{r,s}TM\). Observations by an observer at \(\Gamma(\tau) \in O\) then translate into measurements of the components of a horizontal tensor field with respect to a basis of the corresponding horizontal tangent space \(H_{\Gamma(\tau)}O\), which can conveniently be expressed using the vector fields \(\underline{\tilde{e}}_i\). Finally, also a translation of the matter action~\eqref{eqn:matteraction}, where \(\Phi\) is viewed as a one-form on \(O\) and the projectors~\eqref{eqn:tangbundproj} are used, is straightforward. However, we do not pursue this topic here. Instead we will directly move on to the gravitational dynamics in the next section.

\subsection{Gravity}\label{subsec:cartangravity}
As we have already done in the case of Finsler geometry in section~\ref{subsec:finslergravity}, we now focus on the dynamics of the Cartan geometry, which we identify with the gravitational dynamics. Since gravity is conventionally related to the curvature of spacetime, we will first discuss the notion of curvature in Cartan geometry. We will then derive dynamics for Cartan geometry from an action principle and see how this notion of curvature is involved. For this purpose we will consider two different actions, the first being the Finsler gravity action we encountered before and which we now translate into Cartan language, and an action which is explicitly constructed in terms of Cartan geometric objects.

We start our discussion of curvature in Cartan geometry with its textbook definition:
\begin{defi}[Cartan curvature]
The \emph{curvature} of a Cartan geometry \((\tilde{\pi}: P \to M, A)\) modeled on the homogeneous space \(G/H\) is the $\mathfrak{g}$-valued two-form \(F \in \Omega^2(P,\mathfrak{g})\) on \(P\) given by
\begin{equation}\label{eqn:cartancurvature}
F = dA + \frac{1}{2}[A,A]\,.
\end{equation}
\end{defi}
The curvature has a simple interpretation in terms of the fundamental vector fields \(\underline{A}\): it measures the failure of \(\underline{A}: \mathfrak{g} \to \Vect P\) to be a Lie algebra homomorphism. This can be seen from the relation
\begin{equation}\label{eqn:homofailure}
\underline{A}([a,a']) - [\underline{A}(a),\underline{A}(a')] = \underline{A}(F(\underline{A}(a),\underline{A}(a')))\,,
\end{equation}
which can easily be derived from the definition~\eqref{eqn:cartancurvature} by making use of the standard formula
\begin{equation}
d\sigma(X,Y) = X(\sigma(Y)) - Y(\sigma(X)) - \sigma([X,Y])
\end{equation}
for any one-form \(\sigma\) and vector fields \(X,Y\).

From this general definition we now turn our focus to the Cartan geometry on observer space modeled on \(G/K\), which we derived from Finsler geometry in section~\ref{subsec:cartanobs}. In this context the term \([\underline{A}(a),\underline{A}(a')]\) for \(a,a' \in \mathfrak{z}\) in the relation~\eqref{eqn:homofailure} reminds to the Lie bracket of horizontal vector fields \([\delta_a,\delta_b]\) in the definition of the non-linear curvature \(R^c{}_{ab}\) on \(TTM\). Indeed the similar expression on \(P\) given by
\begin{equation}\label{eqn:horicomm}
[\underline{e}_i,\underline{e}_j] = f_i^bf_j^cf_k^d(\delta_cF^a{}_{bd} - \delta_bF^a{}_{cd} + F^e{}_{bd}F^a{}_{ce} - F^e{}_{cd}F^a{}_{be})\bar{\partial}^k_a
\end{equation}
reproduces the components of the non-linear curvature~\eqref{eqn:nonlincurv}, which can equivalently be written in the form
\begin{equation}\label{eqn:nonlincurv2}
R^a{}_{bc} = y^d(\delta_cF^a{}_{bd} - \delta_bF^a{}_{cd} + F^e{}_{bd}F^a{}_{ce} - F^e{}_{cd}F^a{}_{be})\,.
\end{equation}
We can directly apply this result to the Finsler gravity action~\eqref{eqn:finsleraction} on the unit tangent bundle \(\Sigma \subset TM\). Since observer space is simply the connected component of the unit tangent bundle constituted by the future timelike vectors, it is straightforward to consider the restricted action
\begin{equation}\label{eqn:ffaction}
S_{\text{F}} = \int_O\Vol_{\tilde{G}}R^a{}_{ab}y^b
\end{equation}
as a gravity action on \(O\). This action is still written in terms of Finsler geometric objects, which we will now rewrite in terms of Cartan geometry. For the volume form \(\Vol_{\tilde{G}}\) of the Sasaki metric \(\tilde{G}\) we have already found the expression~\eqref{eqn:cartansasakivol}, while for the non-linear curvature coefficients \(R^a{}_{bc}\) we can make use of the Lie bracket~\eqref{eqn:horicomm} of horizontal vector fields on \(P\) together with the relation~\eqref{eqn:nonlincurv2}. In order to reproduce the scalar quantity \(R^a{}_{ab}y^b\) in the Finsler gravity action from this vector field we further apply the boost component \(b^{\alpha}\) of the Cartan connection and contract appropriately, which yields
\begin{equation}
b^{\alpha}([\underline{e}_{\alpha},\underline{e}_0]) = R^a{}_{ab}f_0^b = R^a{}_{ab}y^b\,.
\end{equation}
The last equality follows from the identification of the tangent vector \(y^a\) with the temporal frame component \(f_0^a\). Note that this expression is a scalar on \(P\) which is constant along the fibers of \(\pi: P \to O\), and can thus be viewed as a scalar on \(O\). We thus finally obtain the gravitational action
\begin{equation}\label{eqn:fcaction}
S_{\text{F}} = \int_Ob^{\alpha}([\underline{e}_{\alpha},\underline{e}_0])\Vol_{\tilde{G}}\,,
\end{equation}
which is now fully expressed in terms of Cartan geometry.

Another possible strategy to obtain gravitational dynamics on the observer space Cartan geometry is to start from general relativity, rewrite the Einstein-Hilbert action in terms of the Cartan connection derived from the metric geometry displayed in section~\ref{subsec:cartanintro}, and finally transform the action to an integral over observer space by introducing an appropriate volume form on the fibers of \(\pi': O \to M\). We will follow this procedure for the remainder of this section. The starting point of this derivation is the action given by MacDowell and Mansouri~\cite{MacDowell:1977jt}. In terms of spacetime Cartan geometry it takes the form~\cite{Wise:2006sm}
\begin{equation}\label{eqn:mmaction}
S_{\text{MM}} = \int_M\kappa_{\mathfrak{h}}(\tilde{F}_{\mathfrak{h}} \wedge \tilde{F}_{\mathfrak{h}})\,.
\end{equation}
Here \(\kappa_{\mathfrak{h}}\) is a non-degenerate inner product on \(\mathfrak{h}\). For simplicity we choose
\begin{equation}
\kappa_{\mathfrak{h}}(h,h') = \tr_{\mathfrak{h}}(h,\star h')\,,
\end{equation}
where \(\tr_{\mathfrak{h}}\) is the Killing form on \(\mathfrak{h}\) and \(\star\) denotes a Hodge star operator. In components we can write the Killing form as
\begin{equation}
\tr_{\mathfrak{h}}(h,h') = h^i{}_jh'^j{}_i
\end{equation}
 and the Hodge star operator as
\begin{equation}
(\star h)^i{}_j = \eta^{im}\eta^{ln}\epsilon_{mjkl}h^k{}_n\,.
\end{equation}
The two-form \(F_{\mathfrak{h}}\) is given by the unique decomposition
\begin{equation}
F = F_{\mathfrak{h}} + F_{\mathfrak{z}}
\end{equation}
of the $\mathfrak{g}$-valued Cartan curvature~\eqref{eqn:cartancurvature} into parts with values in \(\mathfrak{h}\) and \(\mathfrak{z}\). Finally, the tilde indicates that we need to lower this two-form \(F_{\mathfrak{h}}\) on the frame bundle \(P\) to a two-form \(\tilde{F}_{\mathfrak{h}}\) on the base manifold \(M\).

We now aim to lift the action~\eqref{eqn:mmaction} to observer space. For this purpose we need to find a suitable volume form on the fibers of \(\pi': O \to M\). Recall from the definition~\eqref{eqn:observerspace} of observer space that these are given by the future unit timelike shells \(S_x\) for \(x \in M\), which are three-dimensional submanifolds of \(TM\). A natural metric on \(S_x\) is thus given by the restriction of the Sasaki metric \(G\) on \(TM\), or equivalently \(\tilde{G}\) on \(O\), to \(S_x\). Using our results from section~\ref{subsec:cartanobs} on the Cartan geometry of observer space we find that the tangent spaces to \(S_x\) are spanned by the vertical vector fields \(\underline{\tilde{b}}^{\alpha}\), so that the Sasaki metric~\eqref{eqn:cartansasaki} restricts to the Euclidean metric \(\delta_{\alpha\beta}\tilde{b}^{\alpha}\otimes\tilde{b}^{\beta}\). Its volume form is given by
\begin{equation}\label{eqn:verticalvolume}
\Vol_S = \epsilon_{\alpha\beta\gamma}\tilde{b}^{\alpha} \wedge \tilde{b}^{\beta} \wedge \tilde{b}^{\gamma}\,.
\end{equation}
In combination with the action~\eqref{eqn:mmaction} lifted to observer space, which means that \(\tilde{F}_{\mathfrak{h}}\) is now regarded as a two-form on \(O\), this yields the action
\begin{equation}\label{eqn:ccaction}
S_{\text{MM}} = \int_O\kappa_{\mathfrak{h}}(\tilde{F}_{\mathfrak{h}} \wedge \tilde{F}_{\mathfrak{h}}) \wedge \Vol_S\,.
\end{equation}
In order to analyze the terms in this action we make use of the algebra relations~\eqref{eqn:algebra} to decompose \(F_{\mathfrak{h}}\) in the form
\begin{equation}
F_\mathfrak{h} = d\omega + \frac{1}{2}[\omega,\omega] + \frac{1}{2}[e,e] = F_{\omega} + \frac{1}{2}[e,e]
\end{equation}
into the curvature \(F_{\omega}\) of \(\omega\) and a purely algebraic term \(\frac{1}{2}[e,e]\). Using the expressions~\eqref{eqn:connectionz} for \(e\) and~\eqref{eqn:connectionh} for \(\omega\) these take the form
\begin{subequations}
\begin{align}
F_{\omega}{}^j{}_i &= -\frac{1}{2}f^{-1}{}^j_df_i^c\left(R^d{}_{cab}dx^a \wedge dx^b + 2P^d{}_{cab}dx^a \wedge \delta f_0^b + S^d{}_{cab}\delta f_0^a \wedge \delta f_0^b\right)\,,\\
[e,e]^j{}_i &= 2f^{-1}{}^j_af^{-1}{}^k_b\eta_{ik}\sgn\Lambda\,dx^a \wedge dx^b\,,
\end{align}
\end{subequations}
where we have introduced the shorthand notation \(\delta f_0^a = df_0^a + N^a{}_bdx^b\). The coefficients \(R^d{}_{cab}\), \(P^d{}_{cab}\) and \(S^d{}_{cab}\) we find here are the coefficients~\eqref{eqn:clcurvcomp} of the curvature of the Cartan linear connection, which is not surprising, since we used the Cartan linear connection in the definition~\eqref{eqn:covderivative2} of \(\omega\). The term \([e,e]\) depends on the choice of the group \(G\), and thus on the sign of the cosmological constant on the underlying homogeneous space. Applying this decomposition to the expression \(\kappa_{\mathfrak{h}}(F_{\mathfrak{h}} \wedge F_{\mathfrak{h}})\) in the action~\eqref{eqn:ccaction} we obtain the following terms:
\begin{itemize}
\item
A cosmological constant term:
\begin{equation}
\frac{1}{4}\tr_{\mathfrak{h}}([e,e] \wedge \star[e,e]) = -(\sgn\Lambda)^2\epsilon_{ijkl}e^i \wedge e^j \wedge e^k \wedge e^l\,.
\end{equation}
\item
A curvature term:
\begin{equation}
\tr_{\mathfrak{h}}([e,e] \wedge \star F_{\omega}) = \frac{1}{6}\sgn\Lambda\,g^{F\,ab}R^c{}_{acb}\epsilon_{ijkl}e^i \wedge e^j \wedge e^k \wedge e^l + (\ldots)\,.
\end{equation}
\item
A Gauss-Bonnet term:
\begin{equation}
\tr_{\mathfrak{h}}(F_{\omega} \wedge \star F_{\omega}) = -\frac{1}{96}R_{abcd}R_{efgh}\epsilon^{abef}\epsilon^{cdgh}\epsilon_{ijkl}e^i \wedge e^j \wedge e^k \wedge e^l + (\ldots)\,.
\end{equation}
\end{itemize}
The ellipsis in the expressions above indicates that we have omitted terms which are not horizontal, i.e., which contain the vertical one-form \(b\). These terms do not contribute to the total action since their wedge product with the vertical volume form~\eqref{eqn:verticalvolume} vanishes. Note the appearance of the common term
\begin{equation}
\epsilon_{ijkl}e^i \wedge e^j \wedge e^k \wedge e^l\,,
\end{equation}
which, when lowered to a four-form on \(O\), combines with the vertical volume form~\eqref{eqn:verticalvolume} to the volume form~\eqref{eqn:cartansasakivol} of the restricted Sasaki metric. The total action thus takes the final form
\begin{equation}\label{eqn:ccaction2}
S_{\text{MM}} = \int_O\Vol_{\tilde{G}}\left(\frac{1}{6}\sgn\Lambda\,g^{F\,ab}R^c{}_{acb} - \frac{1}{96}R_{abcd}R_{efgh}\epsilon^{abef}\epsilon^{cdgh} - (\sgn\Lambda)^2\right)\,.
\end{equation}
From this we see that we obtain an action based on the curvature of the Cartan linear connection, as we have briefly discussed towards the end of section~\ref{subsec:finslergravity}, provided that we have chosen a model space \(G/H\) for which \(\Lambda \neq 0\). We also find that we always obtain a non-zero cosmological constant term. The magnitude of the physical cosmological constant can be adjusted by introducing suitable numerical factors into the algebra relations~\eqref{eqn:algebra}, which corresponds to a rescaling of the basis vectors \(\mathcal{Z}_i\).

\subsection{The role of spacetime}\label{subsec:spacetime}
In the previous sections we have discussed the physics on Finsler spacetimes in the language of Cartan geometry. For this purpose we considered a principal $K$-bundle \(\pi: P \to O\) over observer space \(O\) and equipped it with a Cartan connection \(A\) derived from Finsler geometry. This construction allowed us to reformulate significant aspects of Finsler spacetime in purely Cartan geometric terms: the definition of physical and inertial observers, the split of the tangent bundle \(TO\) into horizontal and vertical components which crucially enters the definition of observables and physical fields, the Sasaki metric and its volume measure on \(O\) and finally the dynamics of gravity. It should be remarked that these formulations can be applied to any Cartan geometry \((\pi: P \to O, A)\) modeled on \(G/K\), since they do not explicitly refer to the underlying Finsler geometry, or even the spacetime manifold \(M\). This observation stipulates the question whether an underlying spacetime geometry is at all required, or may not even exist, at least as a fundamental object. In this final section we will discuss this question.

We first discuss whether and how we can reconstruct the Finsler spacetime \((M,L,F)\) if we are given only its observer space Cartan geometry \((\pi: P \to O, A)\), together with the presumption that an underlying Finsler spacetime exists. Recall from its definition~\eqref{eqn:observerspace} that the observer space \(O\) of a Finsler spacetime is the (disjoint) union of the future unit timelike shells \(S_x\) for all spacetime points \(x \in M\). Every spacetime point \(x\) thus corresponds to a non-empty subset \(S_x\) of \(O\). Reconstructing the spacetime manifold from its observer space therefore amounts to specifying an equivalence relation which decomposes \(O\) into subsets, and to equipping the resulting set of equivalence classes with the structure of a differentiable manifold. This can be done by making use of the vertical distribution \(VO\), which is tangent to the shells \(S_x\) and can be expressed completely in terms of Cartan geometry as the span of the vector fields \(\underline{\tilde{b}}_{\alpha}\) defined in~\eqref{eqn:vectfieldproj}. From our presumption that an underlying spacetime manifold exists it follows that \(VO\) is integrable. The Frobenius theorem then guarantees that \(VO\) can be integrated to a foliation of \(O\), with projection \(\pi': O \to M\) onto its leaf space, and further that \(M\) carries the structure of a differentiable manifold so that \(\pi'\) becomes a smooth submersion.

The aforementioned procedure allows us to reconstruct the spacetime manifold \(M\) from observer space Cartan geometry. If we now aim to reconstruct also its Finsler geometry on \(TM\), we immediately see that this will be possible at most for vectors which lie inside the forward light cones \(C_x\). This comes from the fact that in the construction of the Cartan geometry on \(O\) we used only the Finsler geometry on the shells \(S_x\), which yields the Finsler geometry on \(C_x\) by rescaling and using its homogeneity properties. This means that we cannot reconstruct the Finsler geometry on spacelike or lightlike vectors, and in particular we cannot reconstruct the null structure of a Finsler spacetime.

In order to reconstruct the Finsler function \(F\) on the future light cones we need to reconstruct the embedding \(\sigma: O \to TM\) of observer space into the tangent bundle of the spacetime manifold \(M\). For this purpose we make use of the properties of observer trajectories. Recall that in section~\ref{subsec:finslerpm} we applied the canonical lift~\eqref{eqn:canonicallift} to a curve \(\gamma\) on \(M\) in order to obtain a curve \(\Gamma\) on \(TM\), and concluded that the canonical lifts of observer trajectories on \(O\) are exactly those curves \(\Gamma\) whose horizontal tangent vector components are given by the Reeb vector field~\eqref{eqn:reebvector} in section~\ref{subsec:cartanpm}. We can therefore proceed as follows. For \(o \in O\) we choose an observer trajectory \(\Gamma\) in \(O\) so that \(\Gamma(\tau) = o\). We then project \(\Gamma\) to a curve \(\gamma\) on \(M\) using the projection \(\pi'\). The tangent vector \(\dot{\gamma}(\tau)\), which we identify with \(o\) via the embedding \(\sigma\), is then related to \(\dot{\Gamma}(\tau)\) via the differential \(\pi'_*\). This relation yields the formula
\begin{equation}
\sigma(\Gamma(\tau)) = \frac{d}{d\tau}\pi'(\Gamma(\tau)) = \pi'_*(\dot{\Gamma}(\tau)) = \pi'_*(\mathbf{r}(\Gamma(\tau)))\,,
\end{equation}
where we have used the fact that \(\pi'_*\) isomorphically maps the horizontal tangent space \(H_{\Gamma(\tau)}O\) to \(T_{\gamma(\tau)}M\). The embedding \(\sigma\) is thus simply given by
\begin{equation}
\sigma = \pi'_* \circ \mathbf{r}\,.
\end{equation}
Finally, we obtain the Finsler function on timelike vectors by imposing \(F = 1\) on the image \(\sigma(O) \subset TM\) and the homogeneity~\eqref{eqn:finslerhomo}. Note that \(L\) can be any homogeneous function \(L = F^n\) here, since \(F\) is smooth when restricted to the timelike vectors.

We now turn our focus to a general Cartan geometry \((\pi: P \to O, A)\) modeled on \(G/K\) for which we do not presume the existence of an underlying Finsler geometry or even a spacetime manifold. Indeed the latter will in general not exist, as we can already deduce from the reconstruction of a Finsler spacetime detailed above. There we have seen that spacetime naturally appears as the leaf space of a foliation, which we obtained by integrating the vertical distribution \(VO\) on observer space. This procedure fails if \(VO\) is non-integrable. Further, even if \(VO\) integrates to a foliation of \(O\), this foliation may not be strictly simple, i.e., its leaf space may not carry the structure of a differentiable manifold. This means that only a limited class of observer space Cartan geometries, including those derived from Finsler spacetimes, admit for an underlying spacetime manifold. Further, even if a spacetime exists, it may not be a Finsler spacetime, since the reconstructed metric~\eqref{eqn:cartansasaki} may not be the Sasaki metric induced by Finsler geometry.

The question arises whether we can still assign a meaningful physical interpretation to an observer space Cartan geometry if its vertical distribution is non-integrable, so that there is no underlying spacetime. Since any physical interpretation should be given based on the measurement of dynamical, physical quantities by observers, this amounts to the question whether these can meaningfully be defined on an arbitrary observer space Cartan geometry. We have provided these definitions throughout our discussion of observer space in section~\ref{sec:cartan} of this work. Our findings suggest that the notion of spacetime is not needed as a fundamental ingredient in the definition of physical observations, but rather appears as a derived object for a restricted class of Cartan geometries.

\section*{Acknowledgments}
The author is happy to thank Steffen Gielen, Christian Pfeifer and Derek Wise for their helpful comments and discussions. He gratefully acknowledges the full financial support of the Estonian Research Council through the Postdoctoral Research Grant ERMOS115.

\end{document}